\documentclass[a4paper,11pt]{article}
\pdfoutput=1 
\usepackage{jcappub}                     
\pdfoutput=1 
                     
\usepackage{graphicx}
\usepackage{lscape}
\usepackage{caption}
\usepackage{subcaption}
\usepackage{float}	
\usepackage{amssymb}
\usepackage[T1]{fontenc} 
\usepackage{color}
\usepackage{xcolor}
\usepackage{pagecolor,lipsum}
\usepackage{mdframed}
\usepackage{multirow}
\usepackage{amsmath}
\usepackage{textcomp}
\usepackage{hyperref}
\usepackage{arydshln}
\usepackage{enumerate}
\usepackage{mathrsfs}
\usepackage{verbatim}
\usepackage{slashed}
\usepackage{epsfig}
\usepackage{enumerate}
\usepackage{color}
\usepackage{wrapfig}
\usepackage{amsfonts}
\usepackage{multirow}
\usepackage{tikz}
\usepackage{amsmath}
 \usepackage[normalem]{ulem}

\usepackage{multicol}

 \usepackage[normalem]{ulem}

\newcommand{\ba}{\begin{array}}
\newcommand{\ea}{\end{array}}
\newcommand{\bd}{\begin{displaymath}}
\newcommand{\ed}{\end{displaymath}}
\newcommand{\bsube}{\begin{subequation}}
\newcommand{\esube}{\end{subequation}}
\newcommand{\bea}{\begin{eqnarray}}
\newcommand{\eea}{\end{eqnarray}}
\newcommand{\bal}{\begin{align}}
\newcommand{\ealign}{\end{align}}
\newcommand{\eal}{\end{align}}
\newcommand{\ben}{\begin{enumerate}}
\newcommand{\een}{\end{enumerate}}
\newcommand{\nn}{\nonumber}

\newcommand{\Slash}[1]{{\ooalign{\hfil/\hfil\crcr$#1$}}}

\title{Effective Field Theory approach to lepto-philic self conjugate dark matter}
\author[a,\dagger ]{Hrishabh Bharadwaj \note[${}^\dagger$]{E-mail: hrishabhphysics@gmail.com (corresponding author) }}
\author[a,\$]{ and \, Ashok Goyal \note[${}^{\$}$]{E-mail: agoyal45@yahoo.co.in}}
\affiliation[a]{Department of Physics $\&$ Astrophysics, University of Delhi, Delhi, India.}

\abstract{We study the self conjugate dark matter (DM) particles interacting primarily with the standard model leptons in an effective field theoretical frame work. We consider SM gauge invariant effective contact interactions between the Majorana fermion, real scalar and a real vector DM with leptons by evaluating the Wilson coefficients appropriate for interaction terms upto dimension-8 and obtain constraints on the parameters of the theory from the observed relic density, indirect detection observations and from the DM-electron scattering cross-sections in the direct detection experiments. Low energy LEP data has been used to study sensitivity in the pair production of such low mass $\le$ 80  GeV DM particles. Pair production of DM particles of mass $\ge$ 50 GeV in association with mono-photons at the proposed ILC has rich potential to probe such effective operators.}


\begin{document}
\maketitle

%

\section{Introduction}
\label{intro}
\par Several cosmological and astrophysical observations at the cosmic and galactic scale have pointed towards the existence of dark matter in the Universe. The dark matter constitutes roughly $\sim$ 23\%  of the energy density of the Universe and contributes roughly $\sim$ 75\% of the entire matter existing in the Universe. Planck Collaboration \cite{Aghanim:2018eyx} has measured the dark matter (DM) density to a great precision and has given the relic density value $\Omega_{DM} h^2 = 0.1198 \pm 0.0012$. The nature of the DM has however, remained undetermined so far. Features of DM interactions can be determined from the direct and indirect experiments. The direct detection experiments like DAMA/ LIBRA   \cite{Bernabei:2013xsa, Bernabei:2018yyw}, CoGeNT \cite{Aalseth:2012if}, CRESST   \cite{Angloher:2016rji}, CDMS   \cite{Agnese:2013rvf},
 XENON100   \cite{Aprile:2016swn, Aprile:2017aty}, LUX   \cite{Akerib:2016vxi}
 and PandaX-II   \cite{Cui:2017nnn} are designed to measure the recoil momentum of scattered atom or nucleon by DM in the chemically inert medium of the detector. These experiments of spin-independent (SI) and spin-dependent scattering cross-section in non-relativistic (NR) regime have reached a sensitivity level where $\sigma_{SI}> 8 \times 10^{-47}$ cm$^2$ for DM mass $\sim 30$ GeV. Collider reaches in the present \cite{Hong:2017avi, Kahlhoefer:2017dnp, Mitsou} and proposed \cite{Dreiner:2012xm, Battaglia:2005ie, Rawat:2017fak} colliders aim at identifying the signature of the DM particle production involving mono or di-jet events accompanied by missing energy. So far no experimental observation has made any confirmed detection and as a result a huge DM parameter space has been excluded. The indirect experiments such as FermiLAT   \cite{Ackermann:2015zua, TheFermi-LAT:2015kwa, Fermi-LAT:2016uux}, HESS   \cite{Abramowski:2013ax}, AMS-02   \cite{Aguilar:2014mma, Aguilar:2016kjl} {\it etc.} are looking for the evidence of excess cosmic rays produced in the DM annihilation to Standard Model (SM) particles photons, leptons, $b\  \bar{b}$ and gauge boson pairs etc.
\par Experiments like PAMELA \cite{Adriani:2013uda, Adriani:2008zr} in the last several years have reported an excess in the positron flux without any significant excess in the proton to antiproton flux. The peaks in $e^+ \, e^-$ channel are also observed in ATIC \cite{Panov:2006kf} and PPB-BETS \cite{PPB-BETS} balloon experiments  at around 1 TeV and 500 GeV respectively.  Recently, Dark Matter Particle Explorer (DAMPE) experiment \cite{Ambrosi:2017wek} has also observed a sharp peak around $\sim$ 1.4 TeV favouring the lepto-philic DM annihilation cross-section of the order of $10^{-26}$ cm$^3$/s. The excess in $e^+ \, e^-$  can be either due to astrophysical
 events like high energy emission from the pulsars  or  resulting from DM pair annihilation in our galactic neighborhood preferably to  $e^+\, e^-$ channel.  Since the aforementioned experiments have not  observed any significant excess in anti-proton channel, the DM candidates, if any, appears  to be lepton friendly {\it lepto-philic} and   have  suppressed interaction with quarks at the tree level. 
 \par Most of the effort in understanding the DM phenomenon has revolved around the hypothesis that DM is weakly interacting massive particle (WIMP) with mass lying between several GeV to a few TeV. WIMPs provide the simplest production mechanism for DM relic density from the early Universe. Various UV complete new physics extensions of SM have been proposed essentially to solve the gauge hierarchy problem in the {\it top-down} approach which  include theories like  extra-dimensions 
   \cite{Appelquist:2000nn}, super-symmetry   \cite{Wess:1974tw, Nilles:1983ge, PRoy}, little-Higgs   \cite{Arkani, Cheng}, extended 2-HDM models with singlets as portal of DM interactions \cite{Dutta:2018hcz}   and {\it etc.}   These models naturally provide the DM candidates or WIMPs, whose  mass-scales are close to that of the electro-weak physics. However, the  Direct  detection  experiments  have shrunk the parameter space of  the  simplified  and popular models  where the   WIMPs are made to interact with the visible world via neutral scalars or gauge Bosons.  
\par The model independent DM-SM particle interactions have also been studied in an Effective Field Theory (EFT) approach where the DM-SM interaction mediator is believed to be much heavier than the lighter mass scale of DM and SM interactions. The EFT approach provides a simple, flexible approach to investigate various aspects of DM phenomenology. EFT approach treats the interaction between DM and SM particle as a contact interaction described by non-renormalizable operators. In the context of DM phenomenology, each operator describes different processes like DM annihilation, scattering and DM production in collider searches with each process its own energy scale which is required to be smaller than the cut-off scale $\Lambda_{\rm eff} \gg $ the typical energy $E$. the nature of these interactions is encapsulated in a set of coefficients corresponding to limited number of Lorentz and gauge invariant dimension five and six effective operators constructed with the light degrees of freedom. The constrained parameter space from various experimental data then essentially maps the viable UV complete theoretical models. The generic effective Lagrangian for scalar, pseudo-scalar, vector and axial vector interactions of SM particles with dark matter candidates of spin $0,\ \frac{1}{2}, 1$ and $\frac{3}{2}$ have been studied in the literature \cite{Zheng:2010js, Freitas:2014jla, Savvidy:2012qa, Chang:2017dvm, Dutta:2017jfj, Khojali:2017tuv}.
\par  Sensitivity analysis for DM-quark effective interactions at LHC have been performed \cite{Kahlhoefer:2017dnp, Mitsou,Boveia:2018yeb,CMS:2012bw,Aad:2014wra,Bell:2014tta,Bhattacherjee:2012ch} in a model-independent way for the dominant (a) mono-jet + $\slash\!\!\!\!E_{\rm T}$, (b) mono-$b$ jet + $\slash\!\!\!\!E_{\rm T}$ and (c) mono-$t$ jet + $\slash\!\!\!\!E_{\rm T}$ processes. Similarly, analysis for DM-gauge Boson effective couplings at LHC have been done by the authors in reference \cite{Cotta:2012nj, Chen:2013gya, Crivellin:2015wva}.  The sensitivity analysis of the coefficients for the lepto-philic operators have also been performed through $e^+e^-\to \gamma\ +\ \slash\!\!\!\!E_{\rm T}$ \cite{Chae:2012bq, Chen:2015tia, Fox:2011pm} and $e^+e^-\to Z^0 + \slash\!\!\!\!E_{\rm T}$ \cite{Bell:2012rg, Rawat:2017fak} channels.
\par In the context of deep inelastic lepton-hadron scattering, Gross and Wilczek \cite{Gross:1974cs}  analyzed the twist-2 operators appearing in the operator-product expansion of two weak currents along with the renormalization-group Equations of their coefficients for asymptotically free gauge theories. Similar analysis was done in \cite{Drees:1993bu} for the effective DM - nucleon scattering induced by twist-2 quark operators in the supersymmetric framework where DM is identified with the lightest supersymmetric particle - neutralino. In \cite{Hisano:2010ct,Hisano:2010yh,Hisano:2017jmz} one loop effect in DM-nucleon scattering induced by twist-2 quark and gluonic operators for scalar, vector and fermionic DM particles was calculated.

\par Although there exist many studies of dimension five and six lepto-philic operators, only a few of them are invariant under the SM gauge symmetry. As discussed above, the contribution of the cosmologically constrained effective operators are not only sensitive at DM direct and indirect detection experiments but are also important in direct searches at high energy colliders. In fact the operators which do not meet the SM gauge symmetry requirement, will not be able to maintain the perturbative unitarity \cite{Bell:2015sza} due to their bad high energy behaviour at collider accessible energies comparable to the electroweak scale $\sim 246$ GeV. Thus the remaining dimension five and six operators based on SM gauge symmetry and on the principle of perturbative unitarity may not contribute to $2 \to 2$ scattering processes relevant for direct detection experiments and showed not be considered in production channels at high energy colliders. It is in this context that study of additional SM gauge invariant operators of dimension greater than six is important and needs to be undertaken \cite{Bruggisser:2016ixa,Bruggisser:2016nzw}.

\par In this paper we consider DM current that couples primarily to the SM leptons through the $SU(2)_L \times U(1)_Y$ gauge invariant effective operators. To ensure the invariance of SM gauge symmetry at all energy scales, we restrict our dark matter candidates to be self conjugate : a Majorana fermion, a real spin 0 or a real spin 1 SM gauge singlet. In section 2, we formulate the effective interaction Lagrangian for fermionic, scalar and vector DM with SM leptons via twist-2 dimension eight operators. In section 3, the coefficients of the effective Lagrangian are constrained from the observed relic density and perform a consistency check from indirect and direct experiments. The constraints from the LEP and the sensitivity analysis of the coefficients of the effective operators at the proposed ILC are discussed in section 4. We summarise our results in section 5. 


\section{Effective lepto-philic  DM interactions}
\label{sec:eff_int} 
Following earlier authors \cite{ Hisano:2015bma, Hisano:2011um, Hisano:2015rsa} the interaction between the dark matter particles ($\chi^0,\ \phi^0\ \&\ V^0$) with the standard model leptons is assumed to be mediated by a heavy mediator which can be a scalar, vector or a fermion. The effective contact interaction between the dark matter particles and leptons is obtained by evaluating the Wilson coefficients appropriate for the contact interaction terms upto dimension-8. The mediator mass is assumed to be greater than all the other masses in the model and sets the cut-off scale $\Lambda_{\rm eff}$. We then obtain the following effective operators for self conjugate spin-$\frac{1}{2}$, spin-$0$ and spin-$1$ dark matter particles interacting with the leptons: 

\begin{subequations}
\bea
{\cal L}_{\rm eff.\, Int.}^{\rm spin \,1/2 \,DM} &=& \frac{{\alpha^{\chi^0}_{S}}}{\Lambda_{\rm  eff}^4} {\cal O}_{S}^{1/2}+\frac{{\alpha^{\chi^0}_{T_1}}}{\Lambda_{\rm  eff}^4} {\cal O}_{T_1}^{1/2} +  \frac{{\alpha^{\chi^0}_{AV}}}{\Lambda_{\rm  eff}^2} {\cal O}_{AV}^{1/2}\label{FLag}\\
{\cal L}_{\rm eff.\, Int.}^{\rm spin \,0 \,DM} &=&\frac{{\alpha^{\phi^0}_{S}}}{\Lambda_{\rm  eff}^4} {\cal O}_{S}^{0}+ \frac{{\alpha^{\phi^0}_{T_2}}}{\Lambda_{\rm  eff}^4} {\cal O}_{T_2}^{0} \label{SLag}\\
{\cal L}_{\rm eff.\, Int.}^{\rm spin \,1 \,DM} &=&\frac{{\alpha^{V^0}_{S}}}{\Lambda_{\rm  eff}^4} {\cal O}_{S}^{1} + \frac{{\alpha^{V^0}_{T_2}}}{\Lambda_{\rm  eff}^4} {\cal O}_{T_2}^{1} + \frac{{\alpha^{V^0}_{AV}}}{\Lambda_{\rm  eff}^2} {\cal O}_{AV}^{1} \label{VLag}
\eea
with
\bea
 \mathcal{O}^{1/2}_S&\equiv&m_{\chi^0}\ \left(\bar{\chi^0}\,\chi^0~\right) \,\, m_l\,\, \left(\overline{l}\,l\right)\label{Op_FS}\\ 
  \mathcal{ O}^{1/2}_{T_1}&\equiv& 
\bar{\chi^0}\,i\,\partial^\mu \,\gamma^\nu\, \chi^0 \,\,\,{\cal O}^l_{\mu\nu} \ +\ \text{h.c.}\label{Op_T1}\\
 \mathcal{O}^{1/2}_{\rm AV}&\equiv&\bar{\chi^0}\,\gamma_\mu\,\gamma_5\,\chi^0 \,\,\, \left(\overline{l}\,\gamma^\mu\,\gamma_5\,l\right)\label{Op_FAV}\\
 \mathcal{O}^{0}_S&\equiv&  m_{\phi^0}^2 \ {\phi^0}^2 \,\, m_l\,\, \left(\overline{l}\,l\right)~\label{Op_SS}\\
\mathcal{O}^{0}_{T_2}&\equiv& \,\,\phi^0\,\, i\, \partial^\mu \,\,i\,\partial^\nu\,\, \phi^0\,\, \mathcal{O}^{l}_{\mu\nu}\ +\ \text{h.c.} \label{Op_FT2}\\ 
\mathcal{O}^{1}_S&\equiv& m_{V^0}^2\  {V^0}^\mu \, {V^0}_\mu \,\, m_l\,\, \left(\overline{l}\,l\right)\label{Op_VS}\\
\mathcal{O}^{1}_{T_2}&\equiv& \ {V^0}^\rho\,\, i\, \partial^\mu \,\,i\,\partial^\nu\,\, {V^0}_\rho\,\, \mathcal{O}^{l}_{\mu\nu} \ +\ \text{h.c.}\label{Op_VT2}\\ 
\mathcal{O}^{1}_{\rm AV}&\equiv&  i \ \epsilon_{\mu\nu\rho\sigma}\,\, {V^0}^\mu\ i\ \partial^\nu \ {V^0}^\rho\,\,\, \left(\overline{l}\,\gamma^\sigma \, \gamma_5\,l \right) \label{Op_VAV}
\end{eqnarray}
 
 The effective operators given above can be seen to be $SU(2)_L \otimes U(1)_Y$ gauge invariant by noting that the leptonic bilinear terms written in terms of left and right - handed gauge eigen-states $l_L$ and $e_R$ can be combined to give the above operators. The term proportional to the lepton mass $m_l$ is obtained by integrating out the Higgs in the EFT formalism. The validity of this term is however, upto the weak scale.
  
 The twist-2  operators ${\cal O}^{l}_{\mu\nu}$ for charged leptons are defined as   

\bea
\label{twist2op}
 {\cal O}^{l}_{\mu\nu}&\equiv&\frac{i}{2}\ \overline{l_L}\ \biggl(
{D}_\mu^L\gamma_\nu^{} +{{D}_\nu^{L}}\gamma_\mu^{}-\frac{1}{2}g_{\mu\nu}^{}
{\Slash{D}^L}\biggr)\ l_L\ +\ \frac{i}{2}\ \overline{e_R}\ \biggl(
{D}_\mu^R\gamma_\nu^{} +{{D}_\nu^{R}}\gamma_\mu^{}-\frac{1}{2}g_{\mu\nu}^{}
{\Slash{D}^R}\biggr)\ e_R \nn\\
\eea

\noindent where ${D_\mu}^L$ and ${D_\mu}^R$ are the covariant derivatives given by
\bea
D_\mu^L&\equiv& i\ \partial_\mu\ -\ \frac{1}{2}g\ \overrightarrow{\tau}\cdot \overrightarrow{W}_\mu\ +\ \frac{1}{2}g^\prime\ B_\mu \nn\\
D_\mu^R&\equiv& i\ \partial_\mu\ +\ g^\prime\ B_\mu
\eea
\end{subequations}

 The Lorentz structure of the operators determines the nature of dominant DM pair annihilation cross-sections. It turns out that the scalar and the axial-vector operator contributions respectively for fermionic and vector DM are $p$-wave suppressed.

\begin{figure*}[h]
	\centering
\begin{multicols}{2}	
		
	\includegraphics[width=0.49\textwidth,clip]{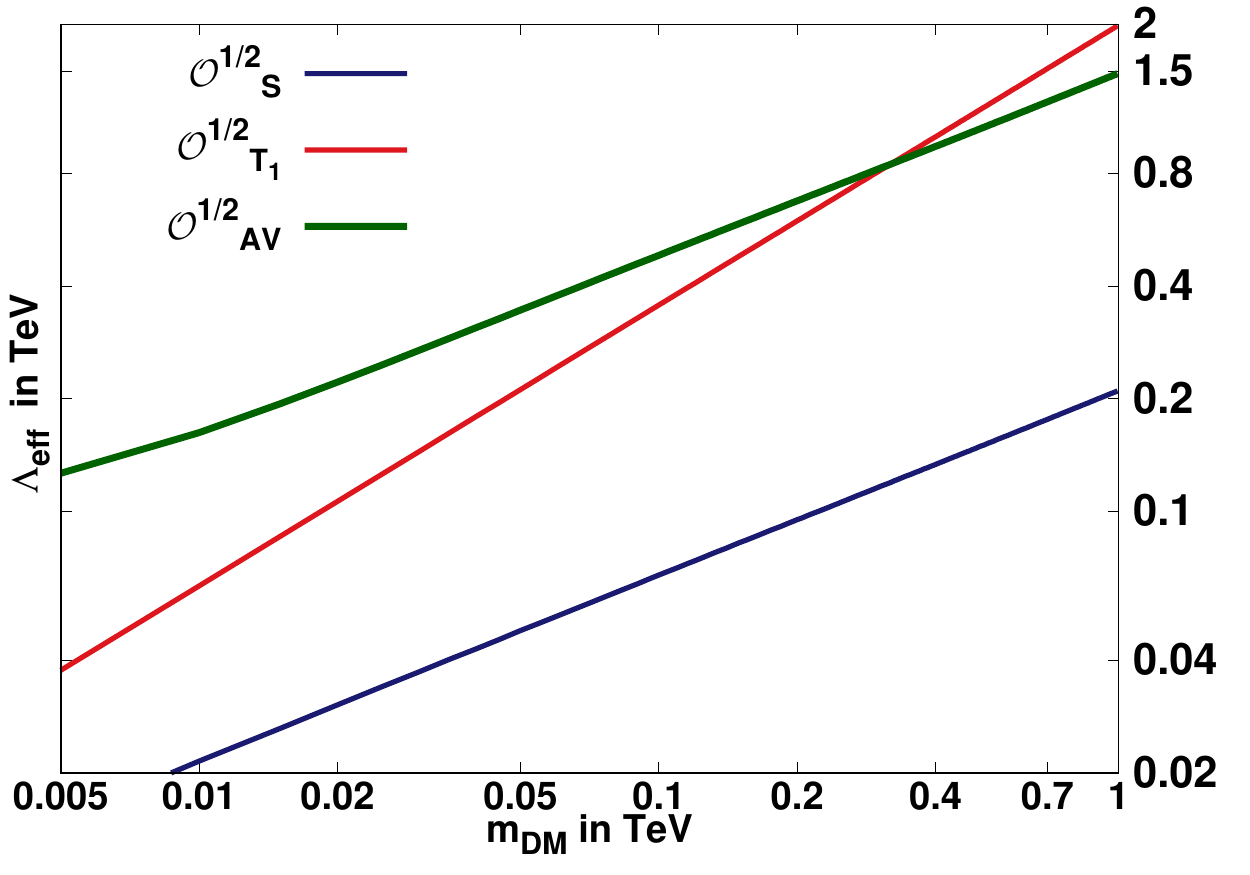} 
	\subcaption{\small \em{ Fermionic DM}}
		\label{lepto-philicrelicdensityF} 
\columnbreak	
	\includegraphics[width=0.49\textwidth,clip]{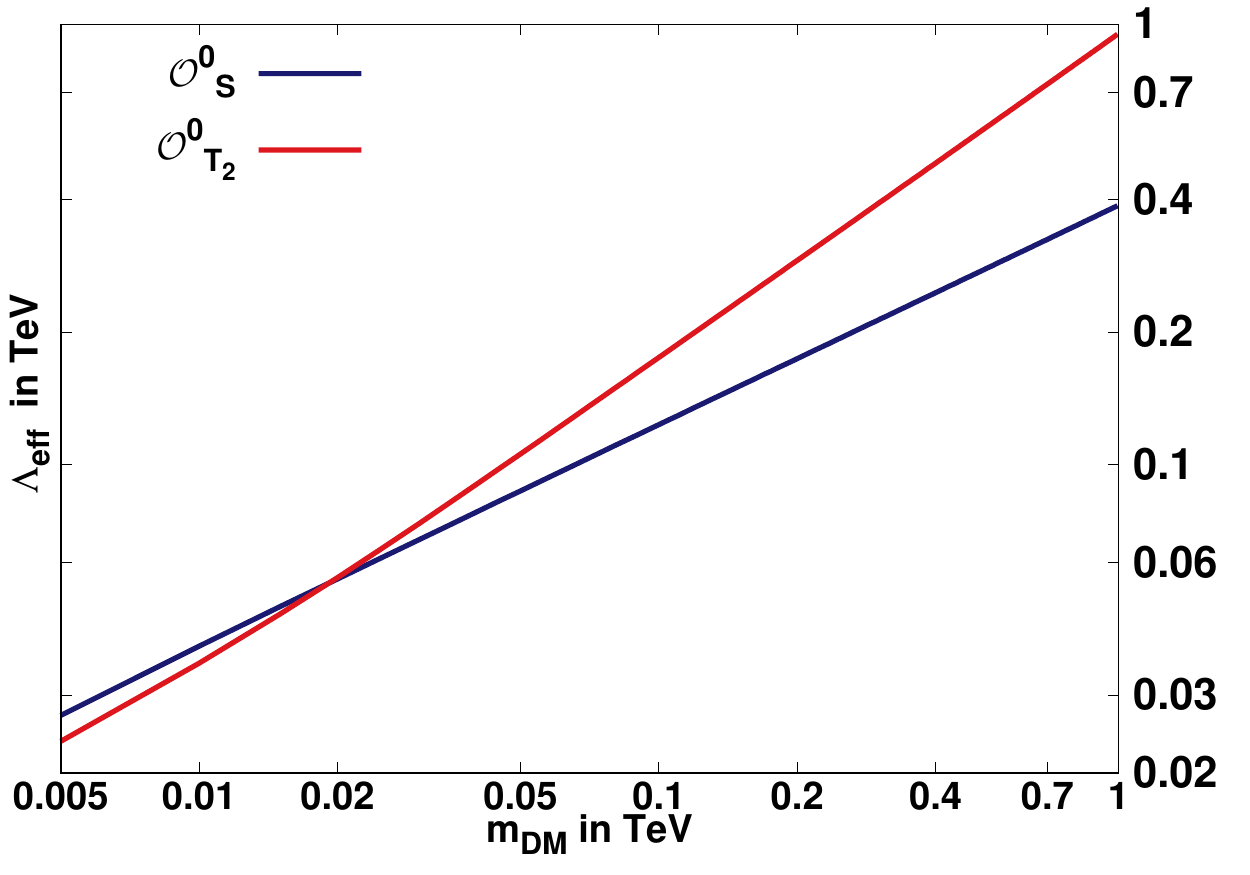}
	\subcaption{\small \em{Scalar DM}}
        \label{lepto-philicrelicdensityS}
\end{multicols}
\begin{center}	
\includegraphics[width=0.5\textwidth,clip]{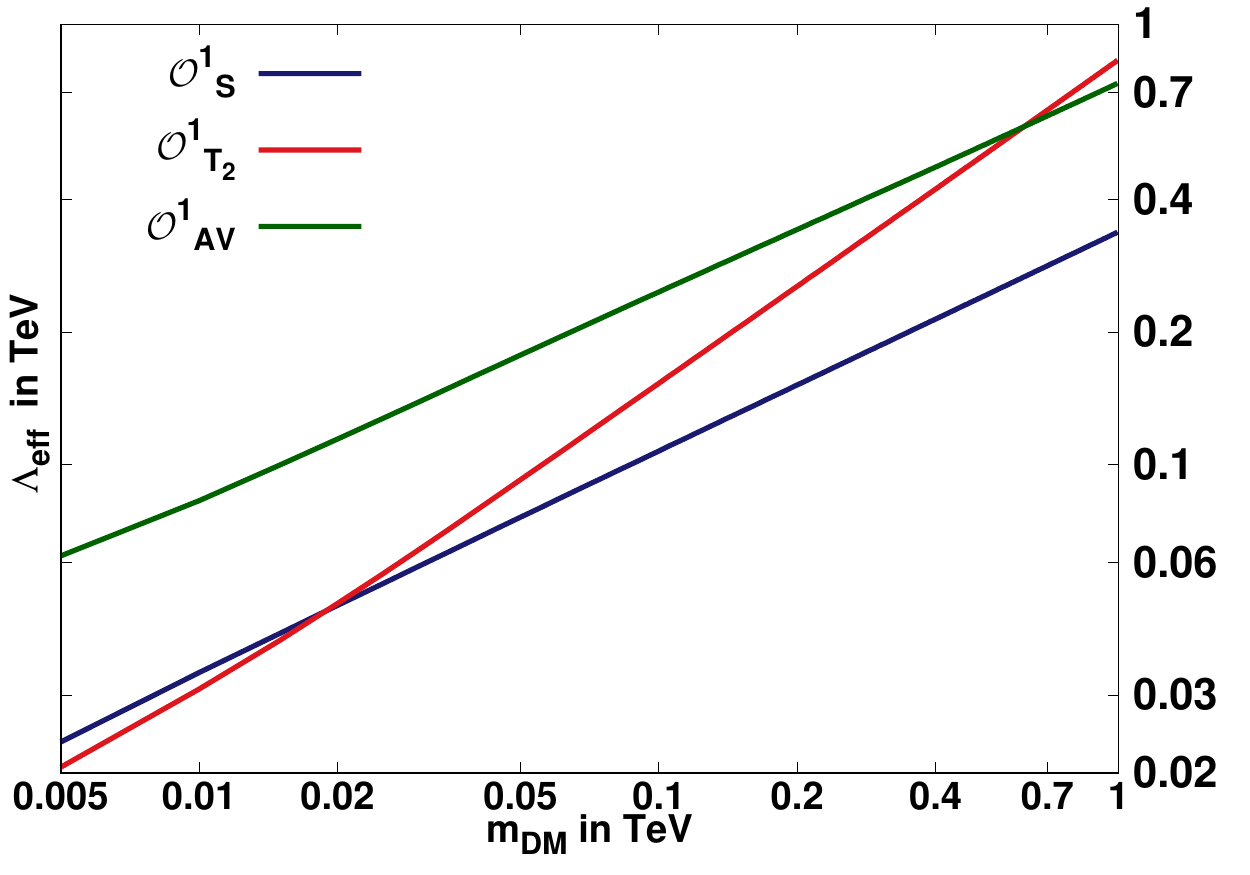}
\subcaption{\small \em{Vector DM}}
\label{lepto-philicrelicdensityV}
\end{center}

	\caption{\small \em {Relic density contours satisfying $\Omega_{\rm DM}h^2$ = $0.1198 \pm 0.0012$ in the DM mass - $\Lambda_{\rm eff}$ plane. All contours are drawn assuming universal lepton flavor couplings of effective DM-lepton interactions. The region below the corresponding solid line is the cosmologically allowed parameter region of the respective operator.}}
\label{fig:relicdensity} 
\end{figure*}
\section{DM Phenomenology}
\label{sec:DMConstraints}
\subsection{Constraints from Relic Density}
\label{subsec:relic}

In the early Universe the DM particles were in thermal equilibrium with the plasma through the creation and annihilation of DM particles. The relic density contribution of the DM particles is obtained by numerically solving the Boltzmann equation \cite{Kolb:1990vq} to give 
\bea
\Omega_{\rm DM}  { h}^2&=&\frac{\pi\,\sqrt{{g_{\rm eff}}(x_F)}}{\sqrt{90}}\frac{x_F\,T_0^3\ g}{M_{Pl}\,\rho_c\,\langle\sigma^{ann} \left\vert \vec{v}\right\vert\rangle\, { g_{\rm eff}}(x_F)}\nonumber\\
&\approx& 0.12 \, \frac{x_F}{28} \, \frac{\sqrt{ g_{\rm eff}(x_F)}}{10}\, \frac{2\times10^{-26} cm^3/s}{\langle\sigma^{ann}\left\vert \vec  v\right\vert\rangle}
\label{boltzmann7}
\eea
and $x_F$ at freeze-out is given by 
\bea
x_F&=&\log \left[a\,\left(a+2\right)\, \sqrt{\frac{45}{8}}\,\frac{ g \,M_{Pl}\, m_{
\rm DM}\,\langle\sigma^{ann}\left\vert \vec  v\right\vert\rangle}{2\,\pi^3\, \sqrt{x_F\,  g_{\rm eff}(x_F)}} \right]
\eea
\noindent where $a$ is a parameter of the order of one. $g_{\rm eff}$ is the effective number of degrees of freedom and is taken to be $92$ near the freeze-out temperature and $g=2,\ 1\ {\rm and}\ 3$ for fermionic, scalar and vector DM particles respectively.

\par The relevant annihilation cross-sections are given in the Appendix \ref{AnnihilationCrosssection}. We have computed the relic density numerically using MadDM \cite{Ambrogi:2018jqj} and MadGraph \cite{Alwall:2014hca} generating the input model file using the Lagrangian given in equations \eqref{FLag}-\eqref{Op_VAV}. In Fig. \ref{fig:relicdensity} we have shown the contour graphs in the effective cut-off $\Lambda_{\rm eff}$ and DM mass plane for the fermionic, scalar and vector DM particles. For arbitrary values of the coupling $\alpha$'s, the effective cut-off $\Lambda_{\rm eff}$ is obtained by noting that $\Lambda_{\rm eff}$ for scalar and twist-2 tensor operators scales as $\alpha^{1/4}$ whereas for AV operators $\Lambda_{\rm eff}$ scales as $\alpha^{1/2}$. We have shown the graphs by taking one operator at a time and taking the couplings $\alpha's=1$. We have made sure that perturbative unitarity of the EFT is maintained for the entire parameter space scanned in Fig. \ref{fig:relicdensity}. The points lying on the solid lines satisfy the observed relic density $\Omega_{DM} h^2=0.1198$. The region below the corresponding solid line is the cosmologically allowed parameter region of the respective operator. We find from Fig. \ref{lepto-philicrelicdensityF} that the scalar  operator for the fermionic DM is sensitive to the low DM mass.

\begin{figure*}
\centering
\begin{multicols}{2}
\includegraphics[width=0.48\textwidth,clip]{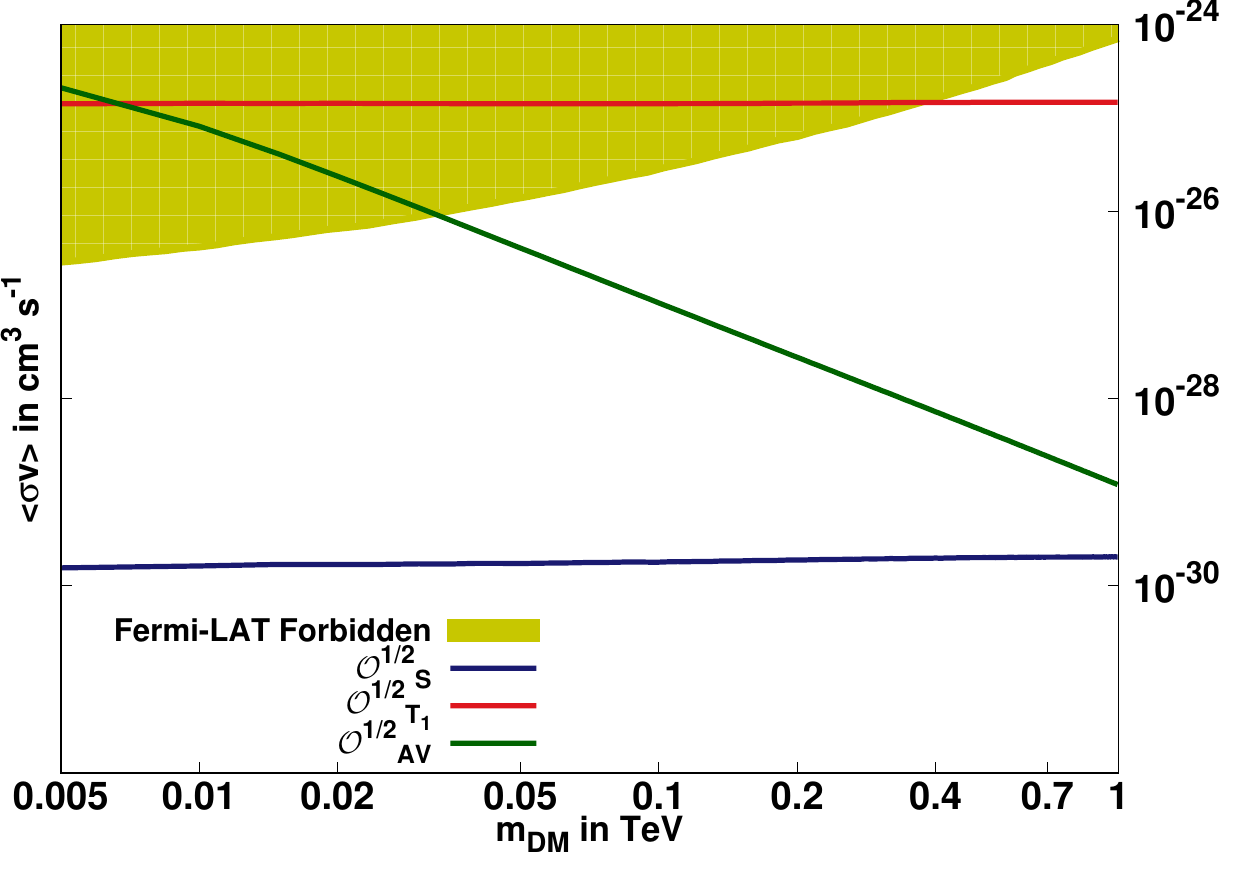}
\subcaption{\small \em{Fermionic DM }}\label{indirectFDM}
\includegraphics[width=0.48\textwidth,clip]{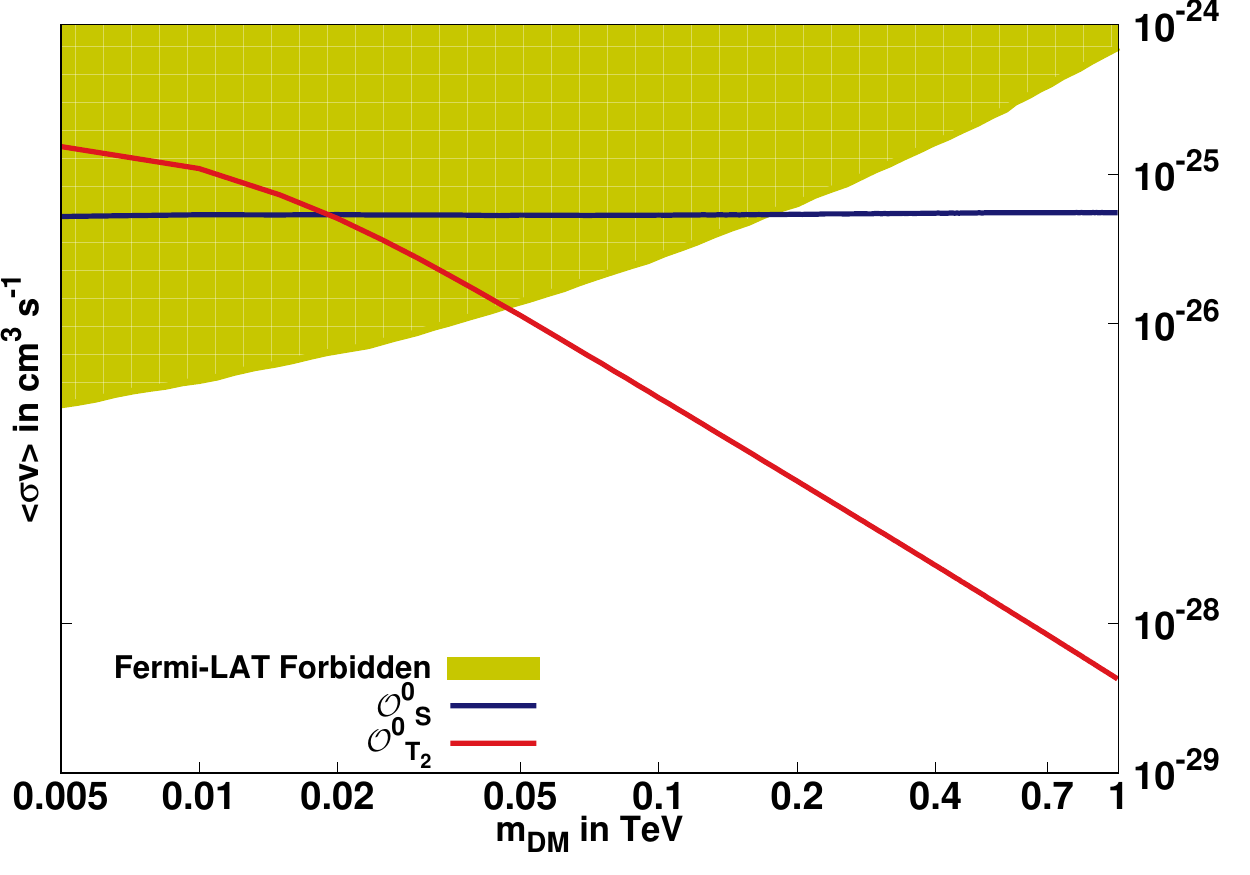}
\subcaption{\small \em{Scalar DM }}\label{indirectSDM}
\end{multicols}
\begin{center}	
\includegraphics[width=0.48\textwidth,clip]{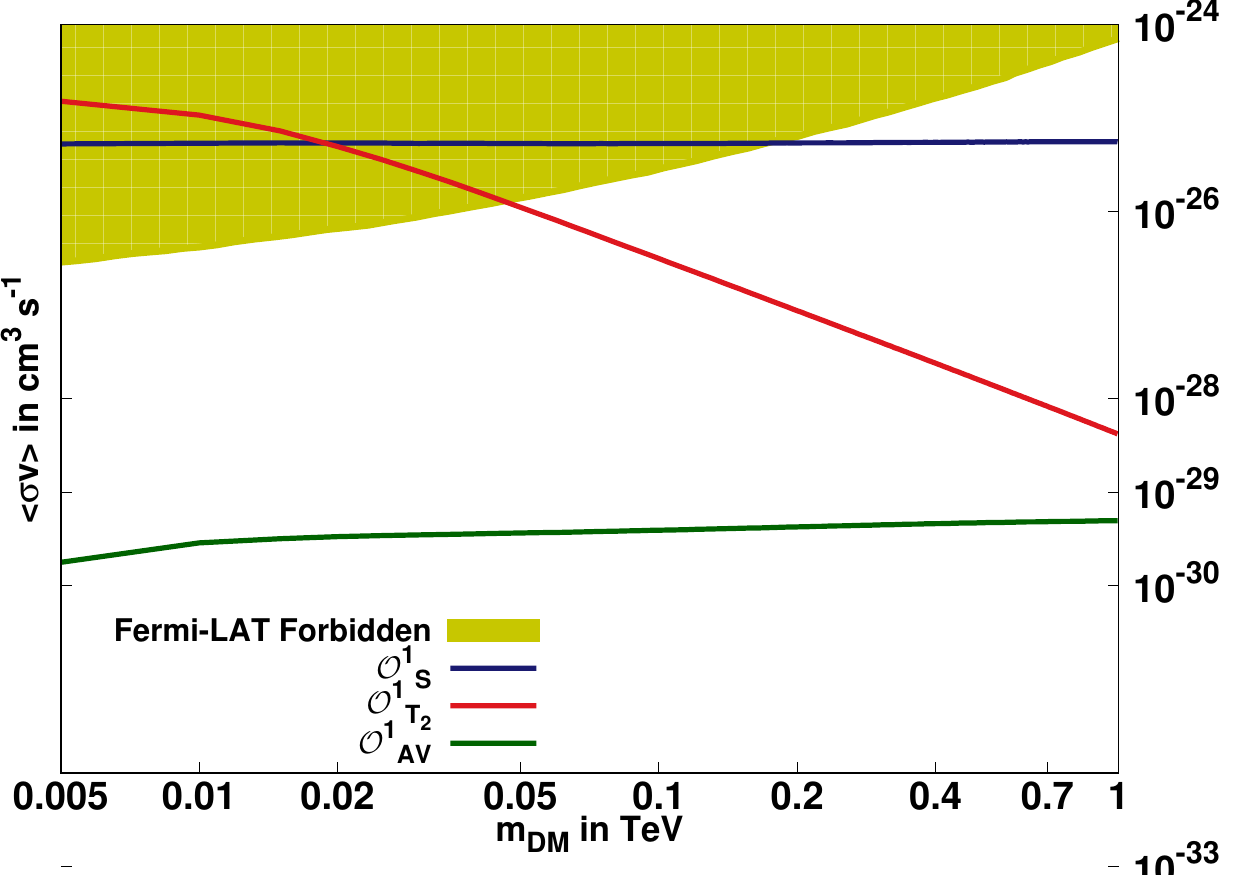}
\subcaption{\small \em{Vector DM }}
\label{indirectVDM}
\end{center}

  \caption{\small \em{ DM annihilation cross-section to $\tau^+ \tau^-$. Solid lines in all figures show the variation of DM annihilation
      cross-section with DM mass where all other parameters are taken from the observed relic density. The median of the DM annihilation cross-section, derived from a combined analysis of the nominal target sample for the  $\tau^+ \tau^-$ channel assuming $100 \%$ branching fraction, restricts the allowed shaded region from above. $v$ is taken to be\ $\sim 10^{-3}\  c$.}}
\label{fig:indirectdetection} 
\end{figure*}
\subsection{Indirect Detection}
\label{subsec:InD}

DM annihilation in the dense regions of the Universe would generate high  flux of the energetic standard model particles. The Fermi Large Area Telescope (LAT) \cite{Ackermann:2015zua, TheFermi-LAT:2015kwa, Fermi-LAT:2016uux} has produced strongest limit on  DM annihilation cross-sections for singular annihilation final states to $b\ \bar{b},\ \tau\ \bar{\tau}$ etc. In the case of DM particles annihilating into multiple channels, the bounds on cross-sections have been analysed in \cite{Carpenter:2016thc}. In our case we display the bounds from Fermi-LAT in Fig. \ref{fig:indirectdetection} by assuming the DM particles considered in this article to couple to only $\tau$-leptons \emph{i.e., } $\tau$-philic DM's.

 In Fig. \ref{fig:indirectdetection} we have shown the prediction for dark matter  annihilation cross-section into $\tau^+ \tau^-$ for the set of parameters which satisfy the relic density constraints for the $\tau$-philic DM particles. These cross-sections are compared with the upper bounds on the allowed annihilation cross-sections in $\tau^+ \tau^-$ channel obtained from the Fermi-LAT data \cite{Ackermann:2015zua, TheFermi-LAT:2015kwa, Fermi-LAT:2016uux}. The Fermi-LAT data puts a lower limit on the DM particle mass even though allowed by the relic-density observations.
  Likewise Fermi-LAT puts severe constraints on the twist-2 ${\cal O}_{T_1}^{1/2}$ operator (Fig. \ref{indirectFDM}) for the fermionic DM and ${\cal O}_{S}^0$ operator (Fig. \ref{indirectSDM}) for the scalar DM. There is a minimum dark matter particle mass allowed by Fermi-LAT observations.

\subsection{DM-electron scattering}
\label{subsec:DDetection}
Direct detection experiments \cite{Bernabei:2013xsa, Bernabei:2018yyw, Aalseth:2012if, Angloher:2016rji, Agnese:2013rvf, Aprile:2016swn, Aprile:2017aty, Akerib:2016vxi, Cui:2017nnn}  look for the scattering of nucleon or atom by DM particles. These experiments are designed to measure the recoil momentum of the nucleons or atoms of the detector material. This  scattering can be broadly classified as (a) DM-nucleon, (b) DM-atom and (c) DM-electron scattering.  Since the lepto-philic DM does not have direct interaction with quarks or gluons at the tree level, the DM-nucleon interaction can only be induced at the loop levels.
\par It has been shown \cite{Kopp:2009et} and has been independently verified by us that the event rate for direct detection of DM-atom scattering is suppressed by a factor of $\sim 10^{-7}$ with respect to the DM-electron elastic scattering which is in turn is suppressed by a factor of $\sim 10^{-10}$ with respect to the loop induced DM-nucleon scattering. In this article we restrict ourselves to the scattering of DM particle with free electrons.
   
\begin{subequations}
\begin{eqnarray}
\sigma_{S}^{\chi^0\, e^-}&=& \frac{{\alpha^{\chi^0}_{S}}^2}{\pi}\ \frac{m_{\chi^0}^2}{\Lambda_{\rm eff}^8}\ m_e^4\ \simeq\ {\alpha^{\chi^0}_{S}}^2\ \left(\frac{m_{\chi^0}}{200\ {\rm GeV}}\right)^2\ \left( \frac{1 {\rm TeV}}{\Lambda_{\rm eff}} \right)^8\ 3.09 \times 10^{-61}\ {\rm cm}^2
\label{DDFSXsec}\\
\sigma_{T_1}^{\chi^0\, e^-}&=&36\ \frac{{\alpha^{\chi^0}_{T_1}}^2}{\pi}\ \frac{m_{\chi^0}^2}{\Lambda_{\rm eff}^8}\ m_e^4\ \simeq\ {\alpha^{\chi^0}_{T_1}}^2\ \left(\frac{m_{\chi^0}}{200\ {\rm GeV}}\right)^2\ \left( \frac{1 {\rm TeV}}{\Lambda_{\rm eff}} \right)^8\ 1.11 \times 10^{-59}\ {\rm cm}^2\nn\\
\label{DDFT1Xec}\\
\sigma_{AV}^{\chi^0\, e^-}&=& 3\  \frac{{\alpha^{\chi^0}_{AV}}^2}{\pi}\ \frac{m_e^2}{\Lambda_{\rm eff}^4}\ \simeq\ {\alpha^{\chi^0}_{AV}}^2\  \left( \frac{1 {\rm TeV}}{\Lambda_{\rm eff}} \right)^4\ 9.27 \times 10^{-47}\ {\rm cm}^2
\label{DDFAVXec}
\end{eqnarray}
\begin{eqnarray}
\sigma_{S}^{\phi^0 \,e^-}&=& \frac{{\alpha^{\phi^0}_{S}}^2}{\pi}\ \frac{m_{\phi^0}^2}{\Lambda_{\rm eff}^8}\ m_e^4\ \simeq\ {\alpha^{\phi^0}_{S}}^2\ \left(\frac{m_{\phi^0}}{200\ {\rm GeV}}\right)^2\ \left( \frac{1 {\rm TeV}}{\Lambda_{\rm eff}} \right)^8\ 3.09 \times 10^{-61}\ {\rm cm}^2
\label{DDSSXsec}\\
\sigma_{T_2}^{\phi^0 \,e^-}&=&\frac{9}{16}\  \frac{{\alpha^{\phi^0}_{T_2}}^2}{\pi}\ \frac{m_{\phi^0}^4}{\Lambda_{\rm eff}^8}\ m_e^2\ \simeq\ {\alpha^{\phi^0}_{T_2}}^2\ \left(\frac{m_{\phi^0}}{200\ {\rm GeV}}\right)^4\ \left( \frac{1 {\rm TeV}}{\Lambda_{\rm eff}} \right)^8\ 2.78 \times 10^{-50}\ {\rm cm}^2\nn\\
\label{DDST2Xsec}
\end{eqnarray}
\begin{eqnarray}
\sigma_{S}^{V^0\, e^-}&=&\frac{{\alpha^{V^0}_{S}}^2}{\pi}\ \frac{m_{V^0}^2}{\Lambda_{\rm eff}^8}\ m_e^4\ \simeq\ {\alpha^{V^0}_{S}}^2\ \left(\frac{m_{V^0}}{200\ {\rm GeV}}\right)^2\ \left( \frac{1 {\rm TeV}}{\Lambda_{\rm eff}} \right)^8\ 3.09 \times 10^{-61}\ {\rm cm}^2
\label{DDVSXsec}\\
\sigma_{T_2}^{V^0\, e^-}&=&\frac{9}{16}\ \frac{{\alpha^{V^0}_{T_2}}^2}{\pi}\ \frac{m_{V^0}^4}{\Lambda_{\rm eff}^8}\ m_e^2\ \simeq\ {\alpha^{V^0}_{T_2}}^2\ \left(\frac{m_{V^0}}{200\ {\rm GeV}}\right)^4\ \left( \frac{1 {\rm TeV}}{\Lambda_{\rm eff}} \right)^8\ 2.78 \times 10^{-50}\ {\rm cm}^2\nn\\
\label{DDVT2Xsec}\\
\sigma_{AV}^{V^0\, e^-}&=&\frac{1}{144}\ \frac{{\alpha^{V^0}_{AV}}^2}{\pi}\ \frac{1}{\Lambda_{\rm eff}^4}\ \frac{m_e^4}{m_{V^0}^2}\ v^4\ \simeq\ {\alpha^{V^0}_{AV}}^2\ \left(\frac{200\ {\rm GeV}}{m_{V^0}}\right)^2\ \left( \frac{1 {\rm TeV}}{\Lambda_{\rm eff}} \right)^4 v^4\  1.34 \times 10^{-60}\ {\rm cm}^2\nn\\
\label{DDVAVXsec}
\end{eqnarray}
\end{subequations}

\par
\begin{figure}
\centering
       \includegraphics[width=0.5\textwidth,clip]{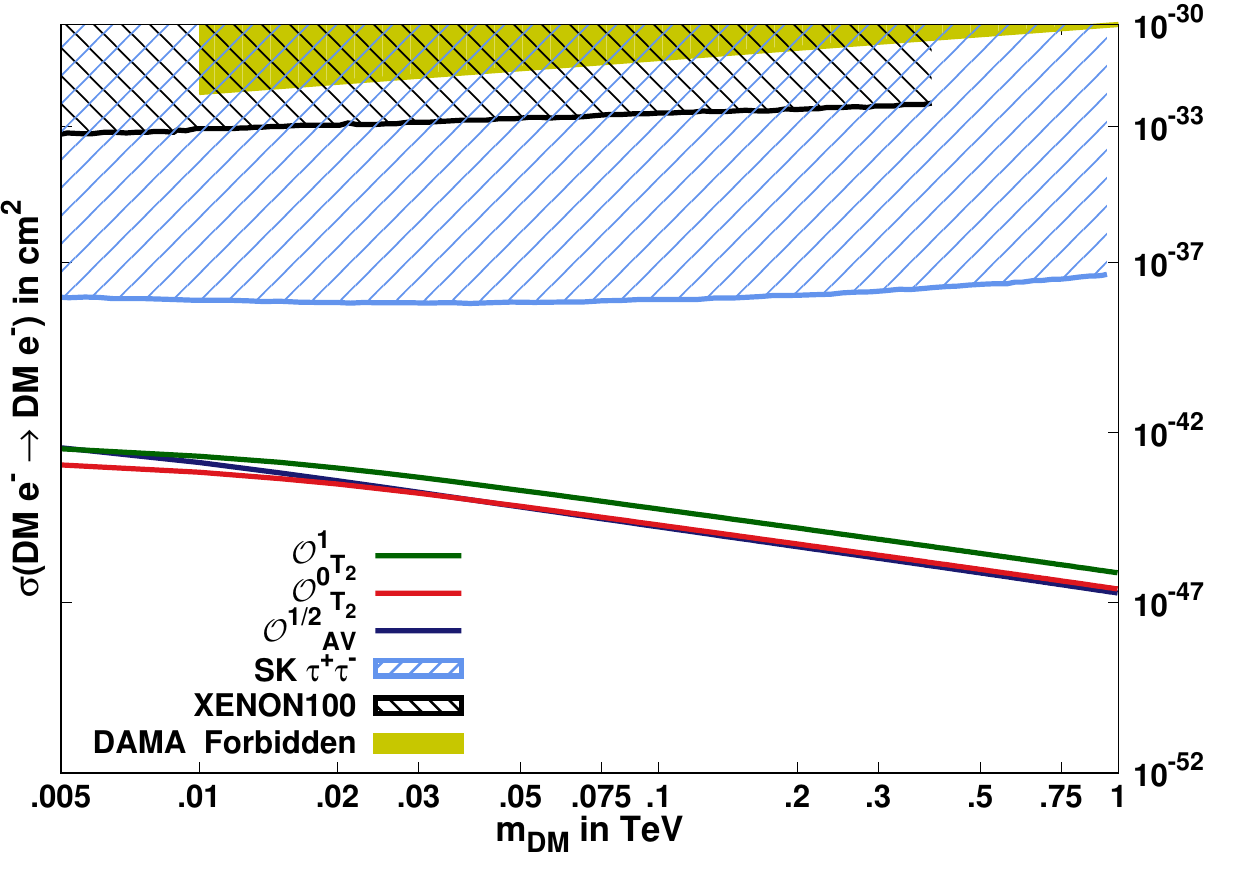}
   \caption{\small \em{DM-free electron elastic scattering cross-section as a function of DM mass. The solid lines are drawn for the dominant operators ${\cal O}_{AV}^{1/2},\ {\cal O}_{T_2}^{0}$ and ${\cal O}_{T_2}^{1}$ for the fermionic, scalar and vector DM particles respectively. The exclusion plots from DAMA at 90\%  C.L. for the case of DM-electron scattering are also shown \cite{Kopp:2009et}. Bounds at 90\% C.L. are shown for XENON100 from inelastic DM-atom scattering \cite{Aprile:2015ade}. The dashed curves show the 90\% C.L. constraint from the Super-Kamiokande limit on neutrinos from the Sun, by assuming annihilation into $\tau^+\tau^-$  \cite{Kopp:2009et}.}}
\label{fig:ddetection} 
\end{figure}

 We find that the electron-DM scattering cross-sections are dominated by the effective interactions mediated by the $AV$ operator ${\cal O}_{AV}^{1/2}$ for the fermionic DM and by the twist-2 operators ${\cal O}_{T_2}^{0}$ and ${\cal O}_{T_2}^{1}$ for the scalar and vector DM respectively. In Fig. \ref{fig:ddetection} we plot the DM-free electron scattering cross-section as a function of DM mass only for the dominant operators as discussed above. The other operators contribution is negligible in comparison. The cross-sections for a given DM mass are computed with the corresponding value of $\Lambda_{\rm eff}$ satisfying the observed relic density for these operators. These results are then compared with the null results of DAMA/LIBRA   \cite{Bernabei:2013xsa, Bernabei:2018yyw}
 at 90\%  confidence level for DM-electron scattering and
 XENON100   \cite{Aprile:2016swn, Aprile:2017aty} at 90\% confidence level for inelastic DM-atom scattering.

\begin{figure*}[tbh]
\centering
\includegraphics[width=0.7\textwidth,clip]{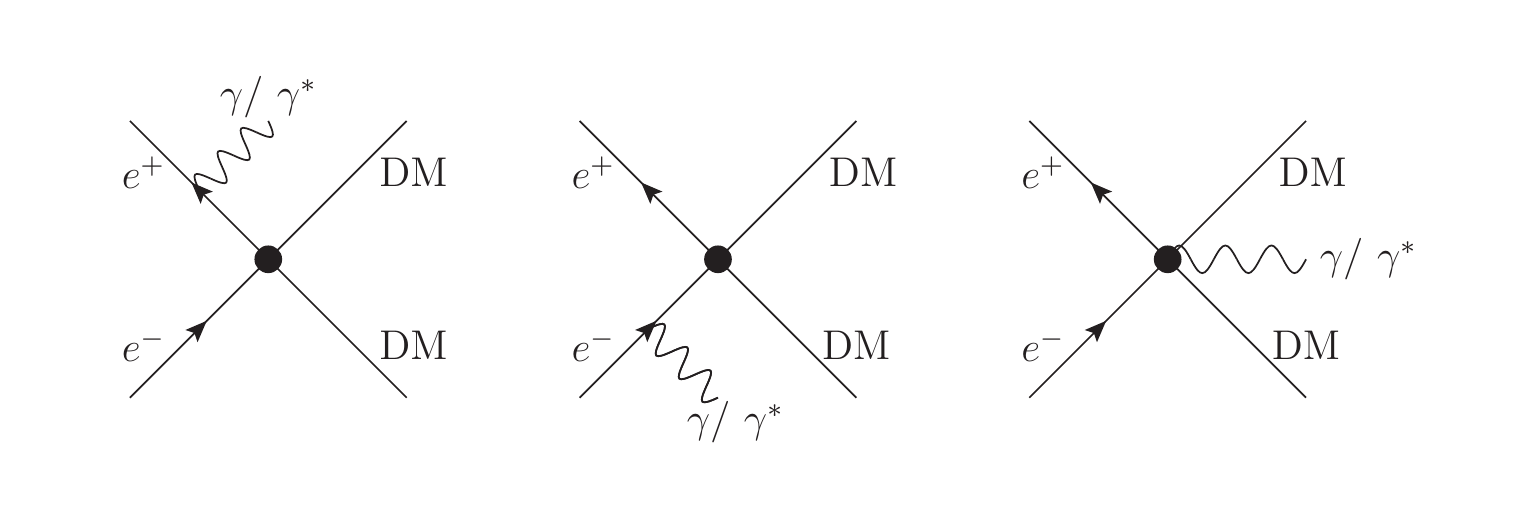}
 
        \caption{\small \em Feynman diagrams contributing to the production of $\gamma / \gamma^\star $ with missing energy induced by lepto-philic operators \eqref{Op_T1}-\eqref{Op_VAV} at the lepton $e^-\,e^+$ collider.}  
        \label{feyndia}
\end{figure*}

\begin{figure*}
\centering
\begin{multicols}{2}
\includegraphics[width=0.49\textwidth,clip]{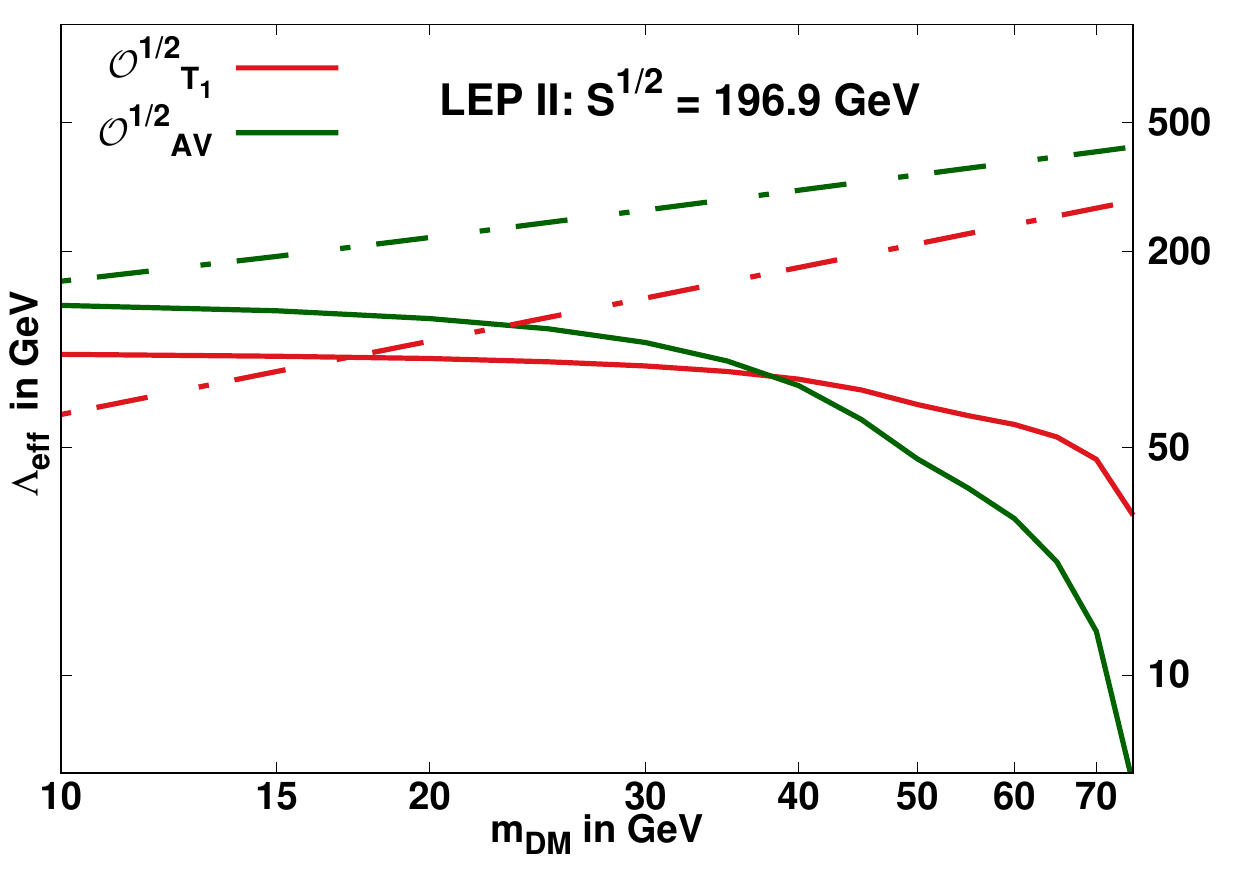}
\subcaption{\small \em{Fermionic DM }}\label{LEPFDM}
\includegraphics[width=0.49\textwidth,clip]{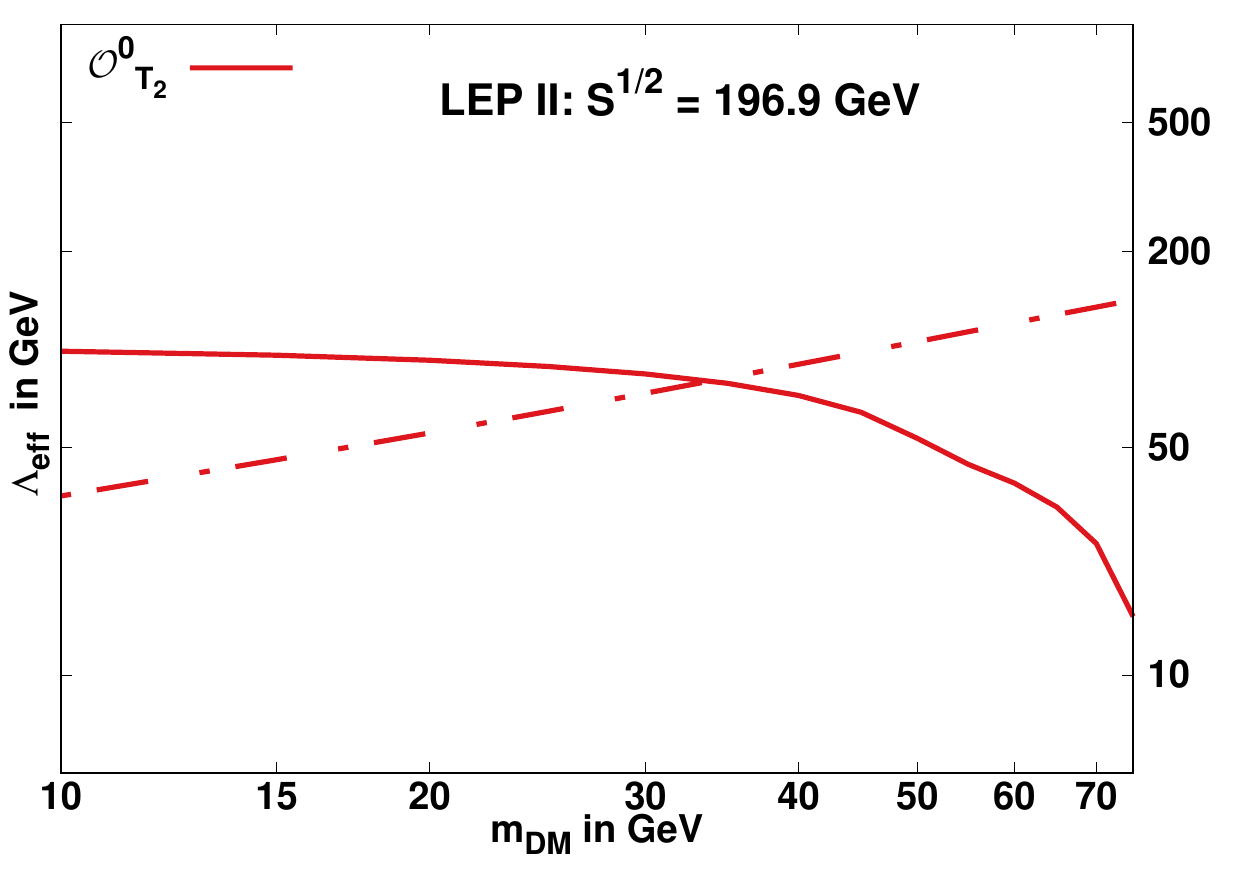}
\subcaption{\small \em{Scalar DM }}\label{LEPSDM}
\end{multicols}
\begin{center}	
\includegraphics[width=0.5\textwidth,clip]{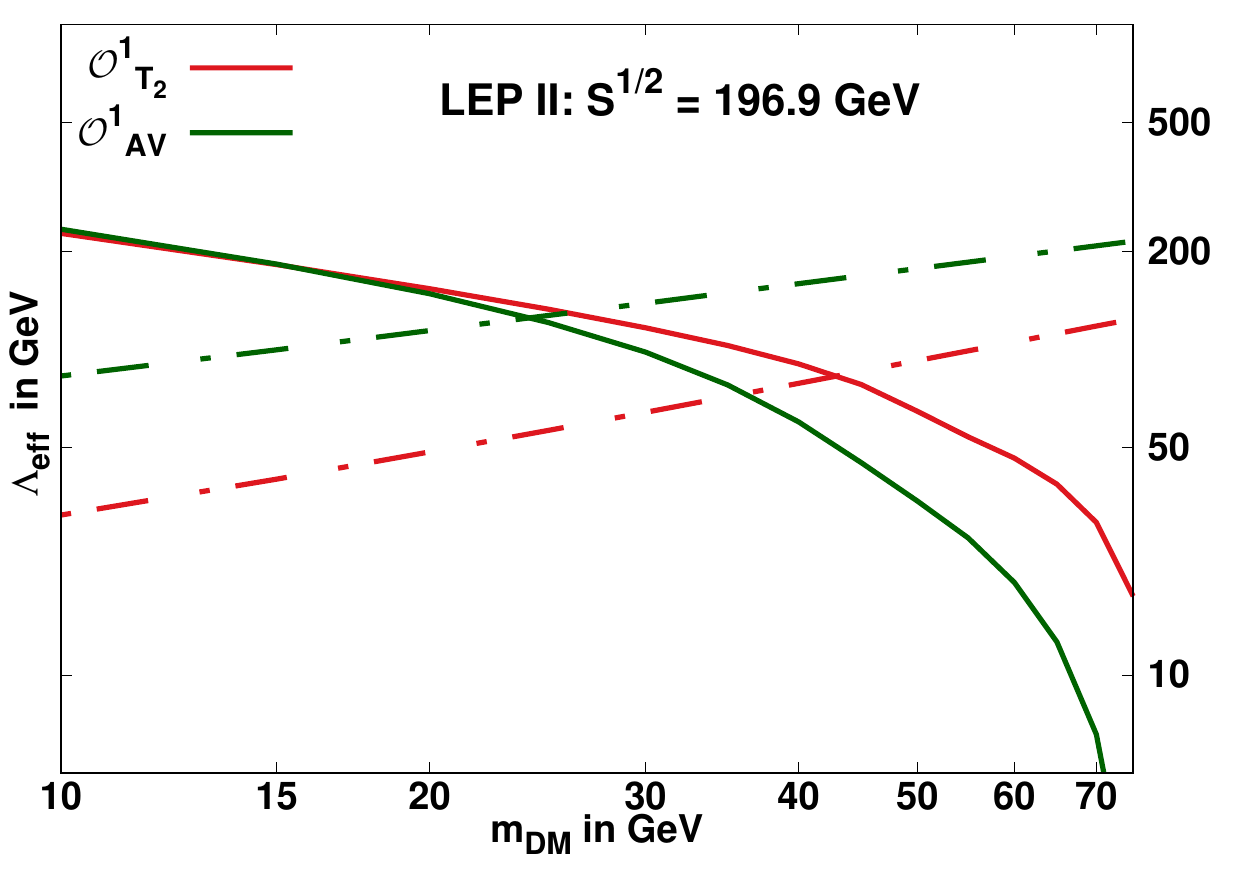}
\subcaption{\small \em{Vector DM }}\label{LEPVDM}
\end{center}

        \caption{\small \em{Solid lines depict the contours in the plane defined by DM mass and the kinematic reach of  for $e^+e^-\to {\rm DM\, pairs} + \gamma^\star \to \,\,\not\!\!\!E_T + q_{i}\bar q_{i} $ at $\sqrt{s}$ = 196.9 GeV and an integrated luminosity of 679.4 pb$^{-1}$, satisfying the constraint $\delta\sigma_{\rm tot}$  = .032 pb obtained from combined analysis of DELPHI and L3 \cite{Schael:2013ita}. The region below solid lines is forbidden by LEP observation. The regions below the dashed lines corresponding to respective operators satisfy  the relic density constraint $\Omega_{\rm DM}h^2 \le$ $0.1198 \pm 0.0012$.}}  
        \label{LEP}
\end{figure*}

\section{Collider sensitivity of effective operators}
\label{sec:collider}
\subsection{LEP Constraints on the effective operators}
\label{subsec:LEPCons}
Existing results and observations from LEP data can be used for putting constraints on the effective operators. The cross-section for the process $e^+e^-\to \gamma^\star + \, {\rm DM \ pair}$ is compared with the combined analysis from DELPHI and L3 collaborations   for $e^+e^-\to \gamma^\star + Z \to q_{i}\bar q_{i} + \nu_{l_j}\bar\nu_{l_j}$ at $\sqrt{s}$ = $196.9$ GeV and an integrated luminosity of 679.4 pb$^{-1}$, where $q_i\equiv u,\,d,\,s$ and $\nu_{l_j}\equiv \nu_e,\,\nu_{\mu},\nu_\tau$. The Feynman diagrams contributing to the production of $\gamma / \gamma^\star $ with missing energy induced by lepto-philic operators at the lepton $e^-\,e^+$ collider are shown in Fig. \ref{feyndia}. The  measured cross-section from the combined analysis for the said process is found to be $0.055$ pb along with the measured statistical error $\delta\sigma_{\rm stat}$, systematic error $\delta\sigma_{\rm syst}$ and total  error $\delta\sigma_{\rm tot}$ of $0.031$ pb, $0.008$ pb and $0.032$ pb respectively \cite{Schael:2013ita}.  Hence, contribution due to an additional channel  containing the final states DM pairs and resulting into the missing energy along with two quark jets can be constrained from the observed $\delta\sigma_{\rm tot}$. In Fig. \ref{LEP} we have plotted the 95\% C.L. solid line contours satisfying  $\delta\sigma_{\rm tot}$$\approx$ $0.032$ pb  corresponding to the operators in the DM mass-$\Lambda_{\rm eff}$ plane. The region under the solid lines corresponding to the operator as shown is disallowed by the combined LEP analysis. The phenomenologically interesting DM mass range $\le 50$ GeV except for the operator ${\cal O}_{AV}^{1/2}$ is completely disfavored by the LEP experiments.

\subsection{$\slash\!\!\! \!E_T$ + Mono-photon signals at ILC and ${{\cal X}}^2$ Analysis}
\begin{figure*}
\centering
\begin{multicols}{2}
\includegraphics[width=0.55\textwidth,clip]{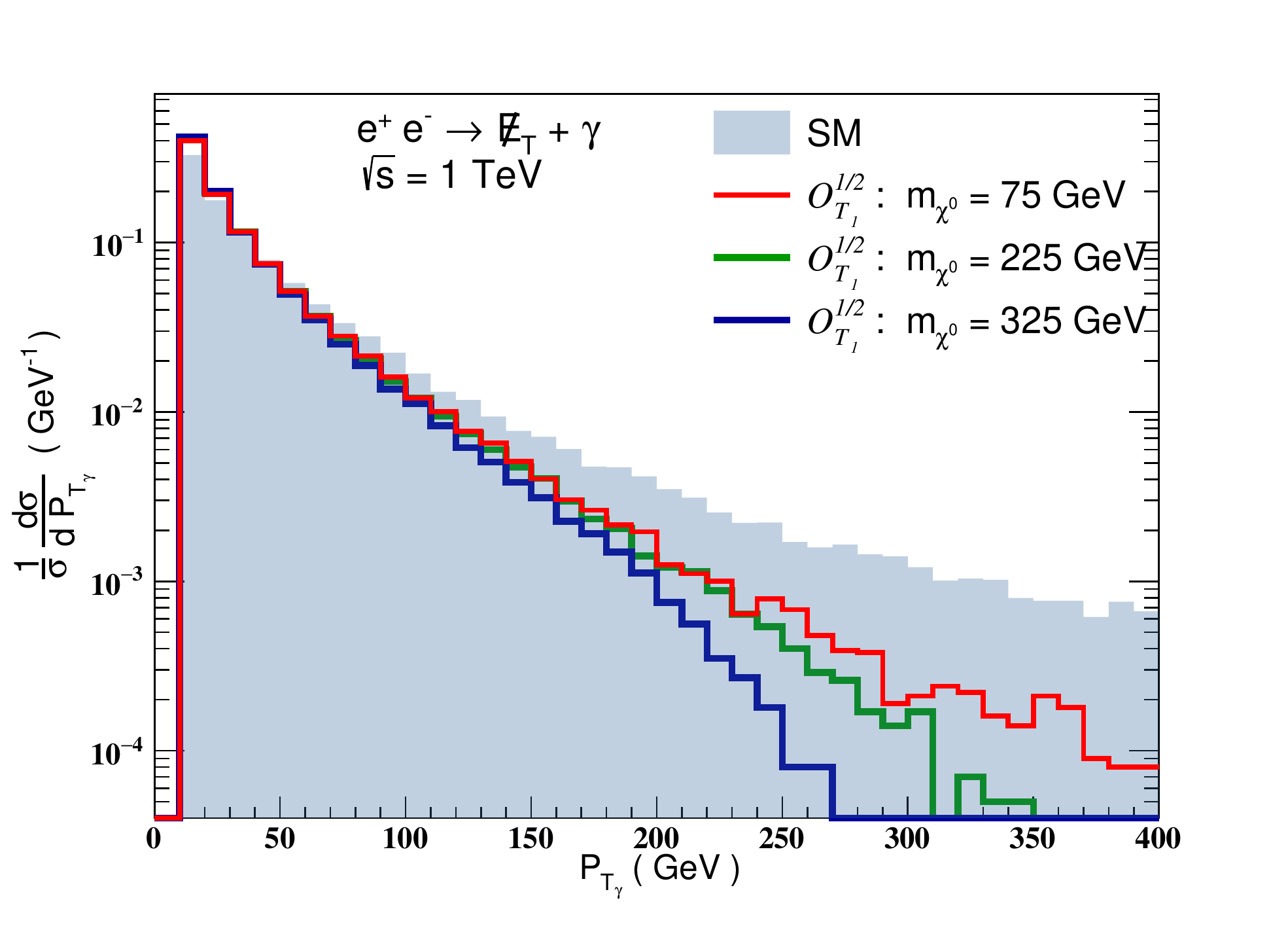}
\subcaption{\small \em{ }}\label{Histo:pT_FT1}
\includegraphics[width=0.55\textwidth,clip]{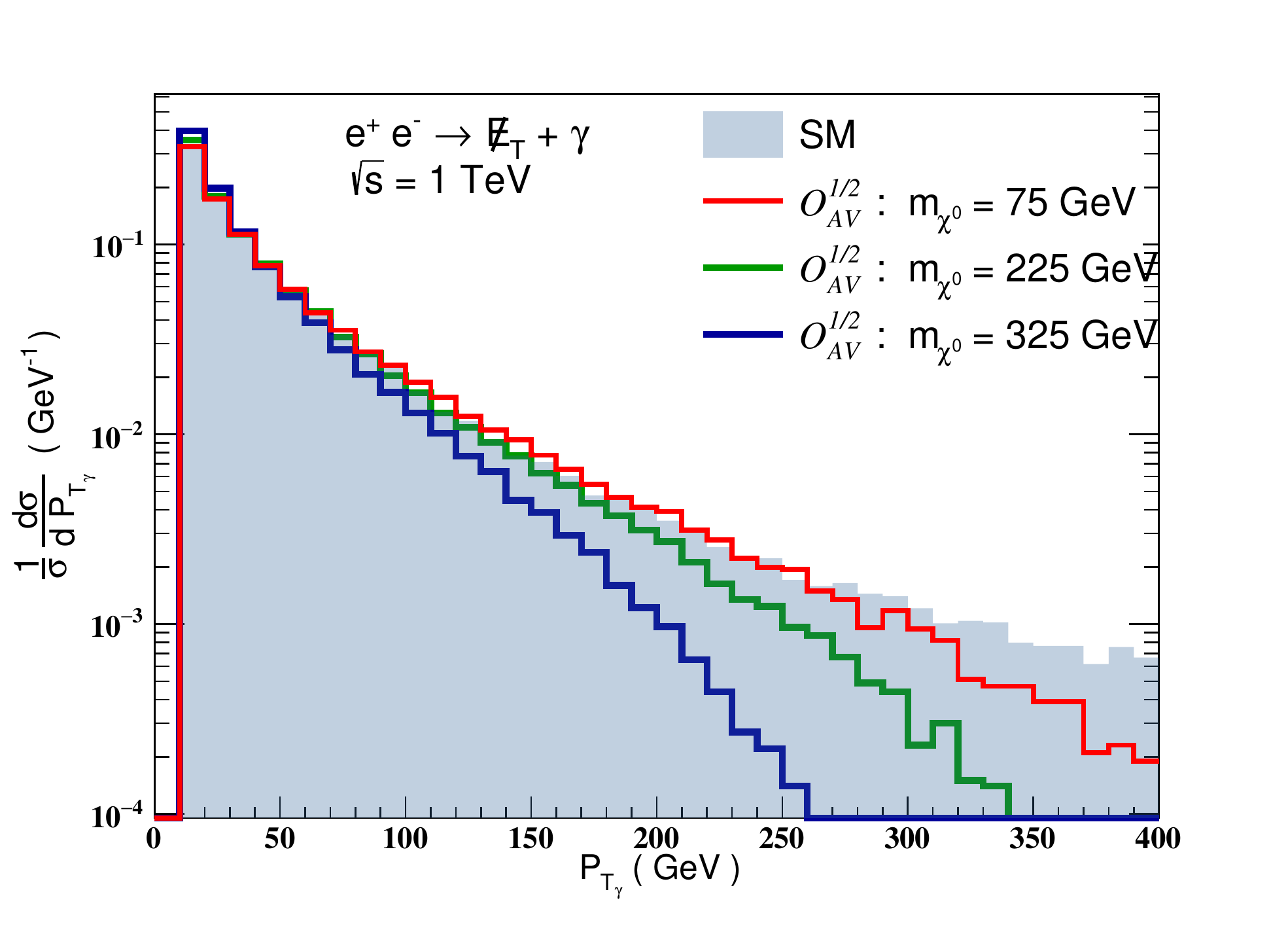}
\subcaption{\small \em{ }}\label{Histo:pT_FAV}
\end{multicols}
\begin{multicols}{2}
\includegraphics[width=0.55\textwidth,clip]{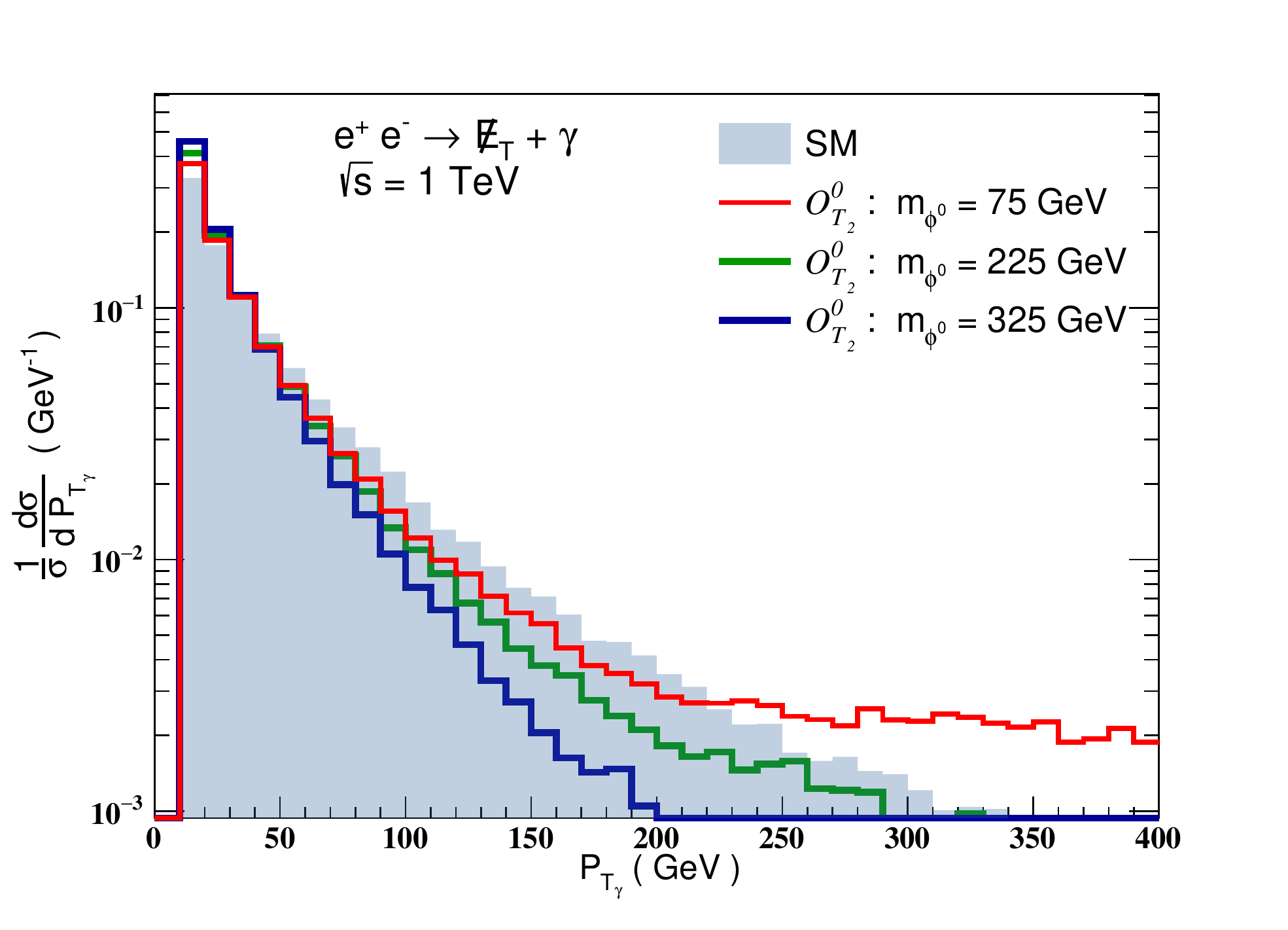}
\subcaption{\small \em{ }}\label{Histo:pT_ST2}
\includegraphics[width=0.55\textwidth,clip]{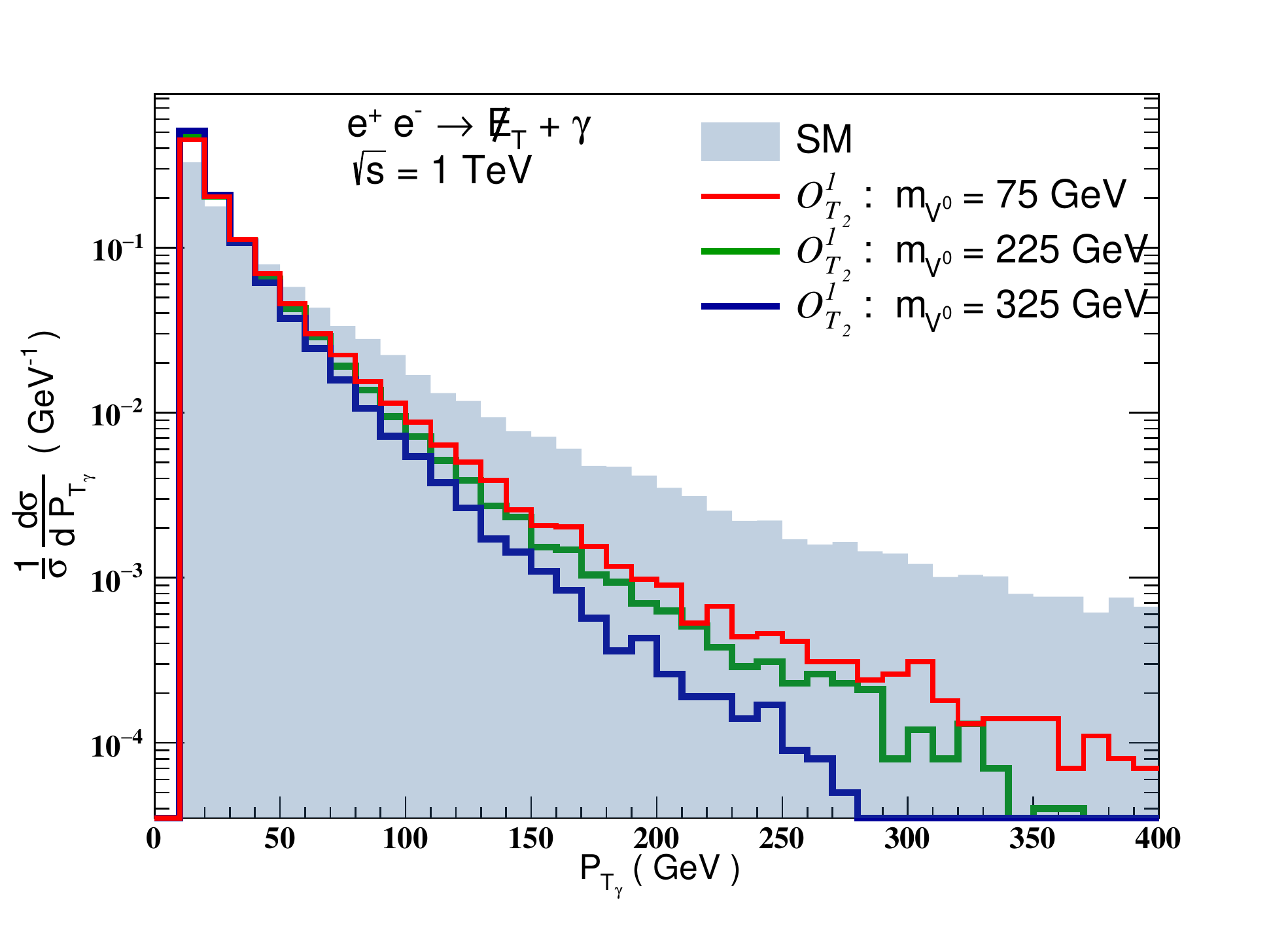}
\subcaption{\small \em{ }}\label{Histo:pT_VT2}
\end{multicols}
\begin{center}	
\includegraphics[width=0.55\textwidth,clip]{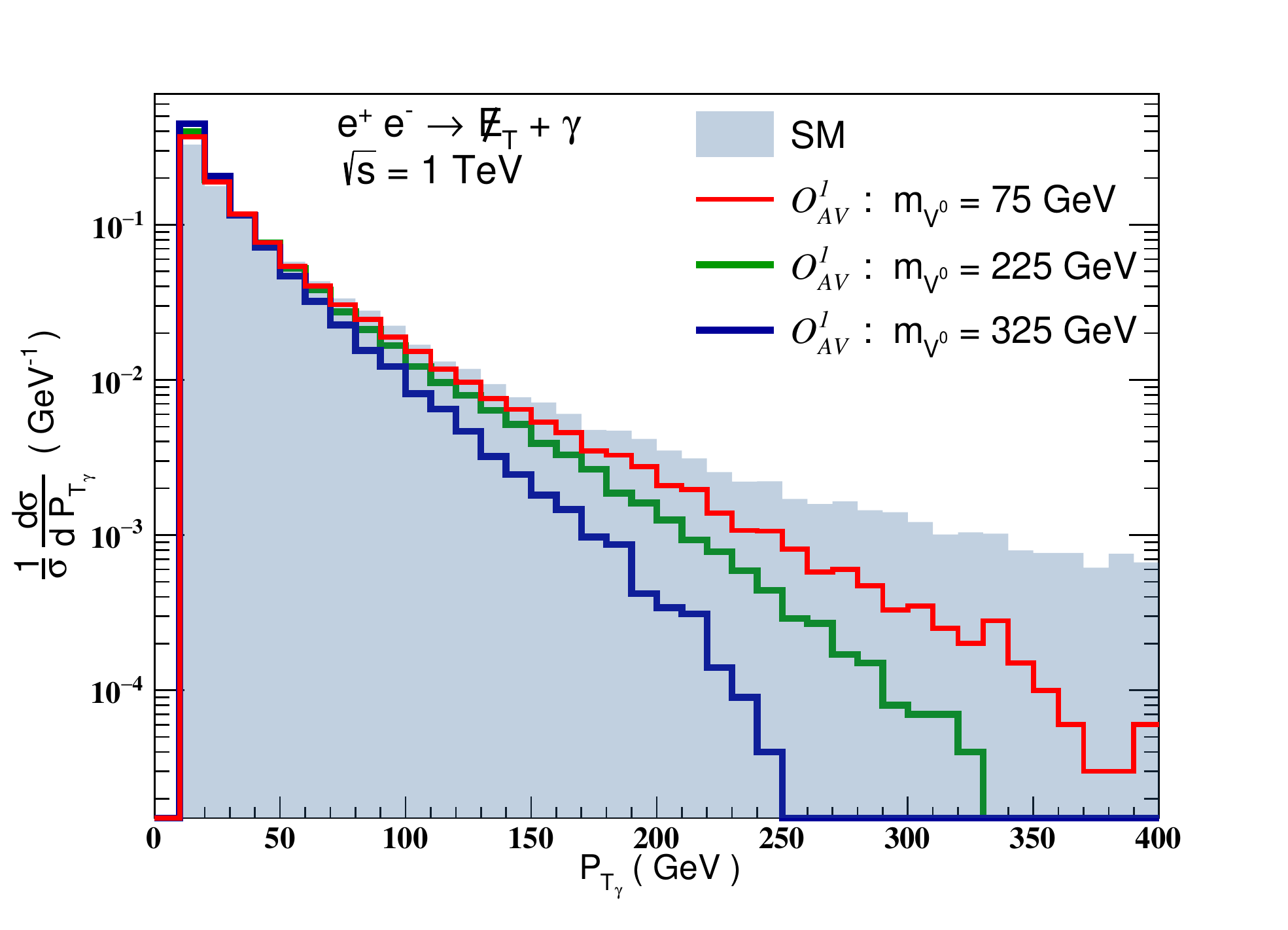}
\subcaption{\small \em{ }}
\label{Histo:pT_VAV}
\end{center}

	\caption{\small \em{Normalized 1-dimensonal differential  cross-sections with respect to $p_{T_\gamma}$ corresponding to the  SM processes and   those induced by  lepto-philic operators  at the three representative values of DM masses: 75, 225 and 325 GeV.}}
\label{fig:distPT}
\end{figure*}

\begin{figure*}
\centering
\begin{multicols}{2}
\includegraphics[width=0.55\textwidth,clip]{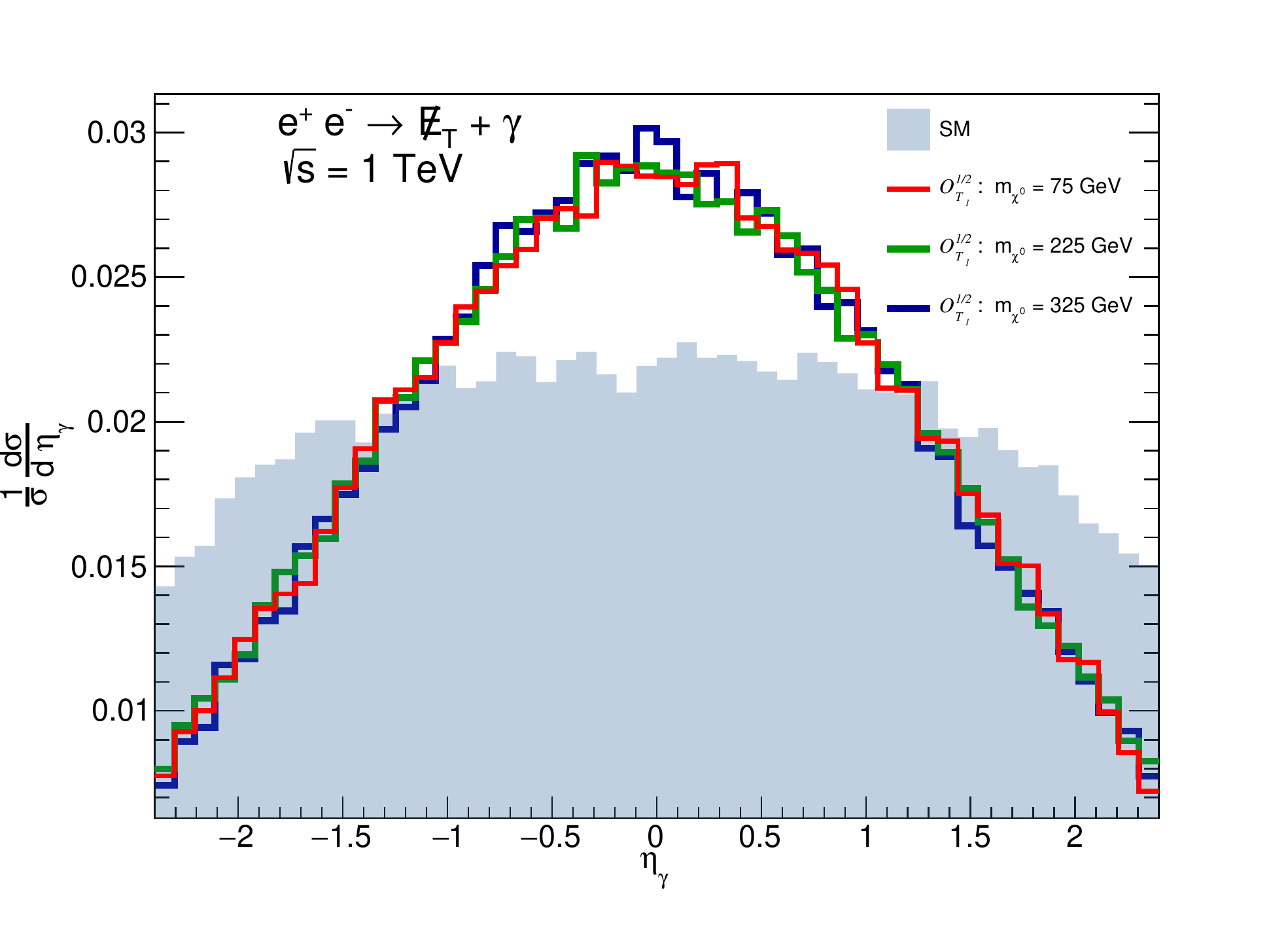}
\subcaption{\small \em{ }}\label{Histo:eta_FT1}
\includegraphics[width=0.55\textwidth,clip]{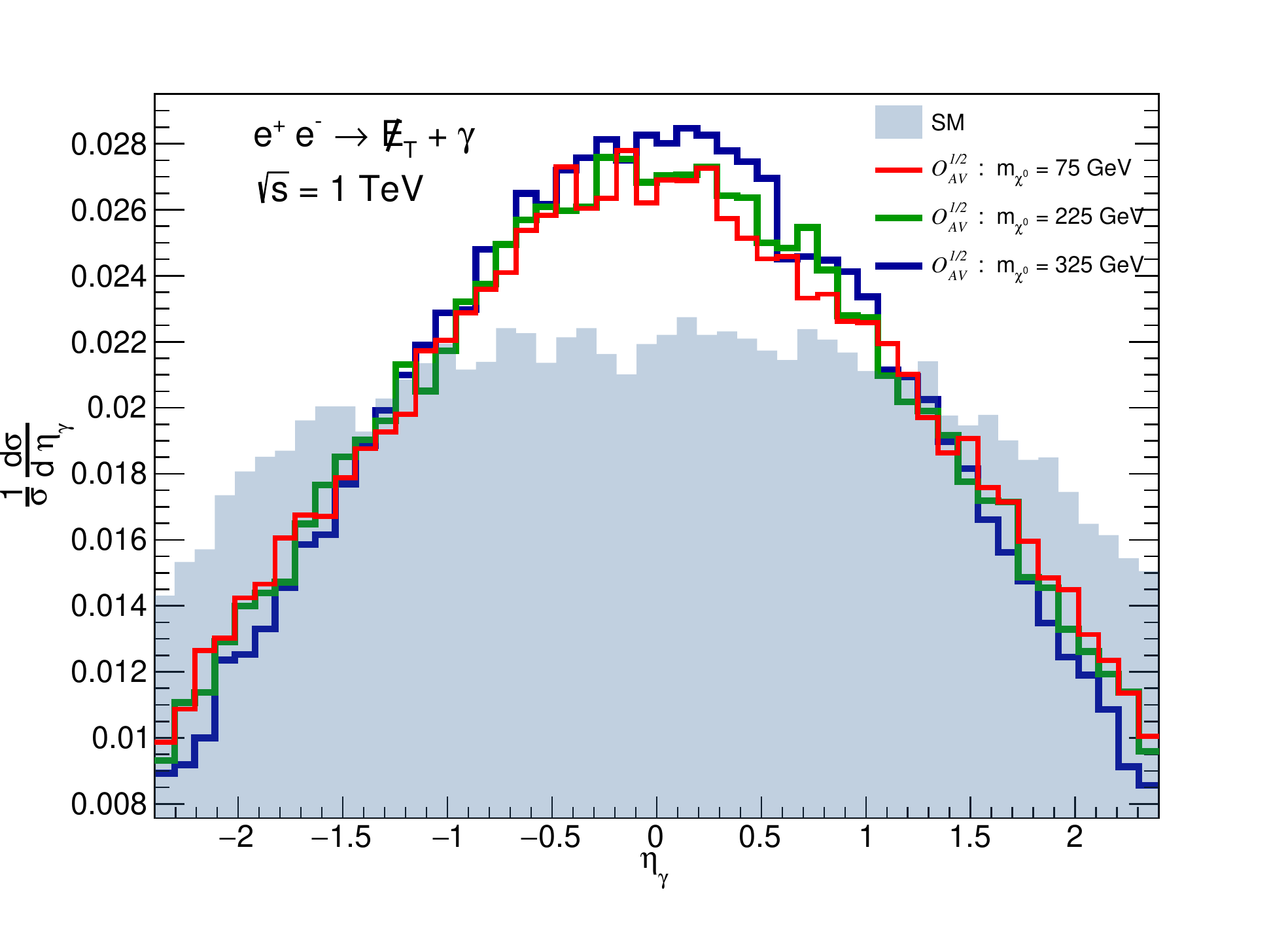}
\subcaption{\small \em{ }}\label{Histo:eta_FAV}
\end{multicols}
\begin{multicols}{2}
\includegraphics[width=0.55\textwidth,clip]{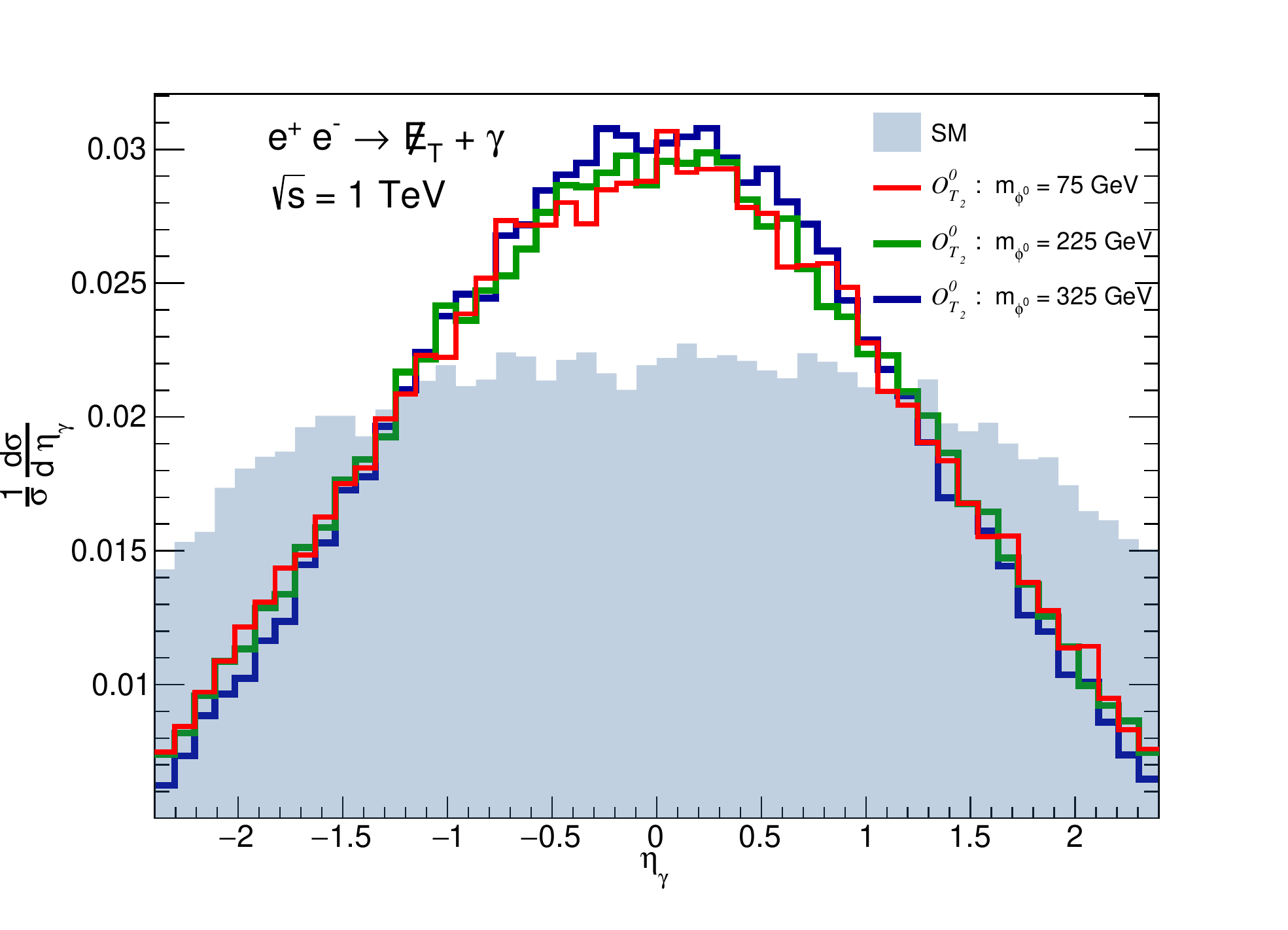}
\subcaption{\small \em{ }}\label{Histo:eta_ST2}
\includegraphics[width=0.55\textwidth,clip]{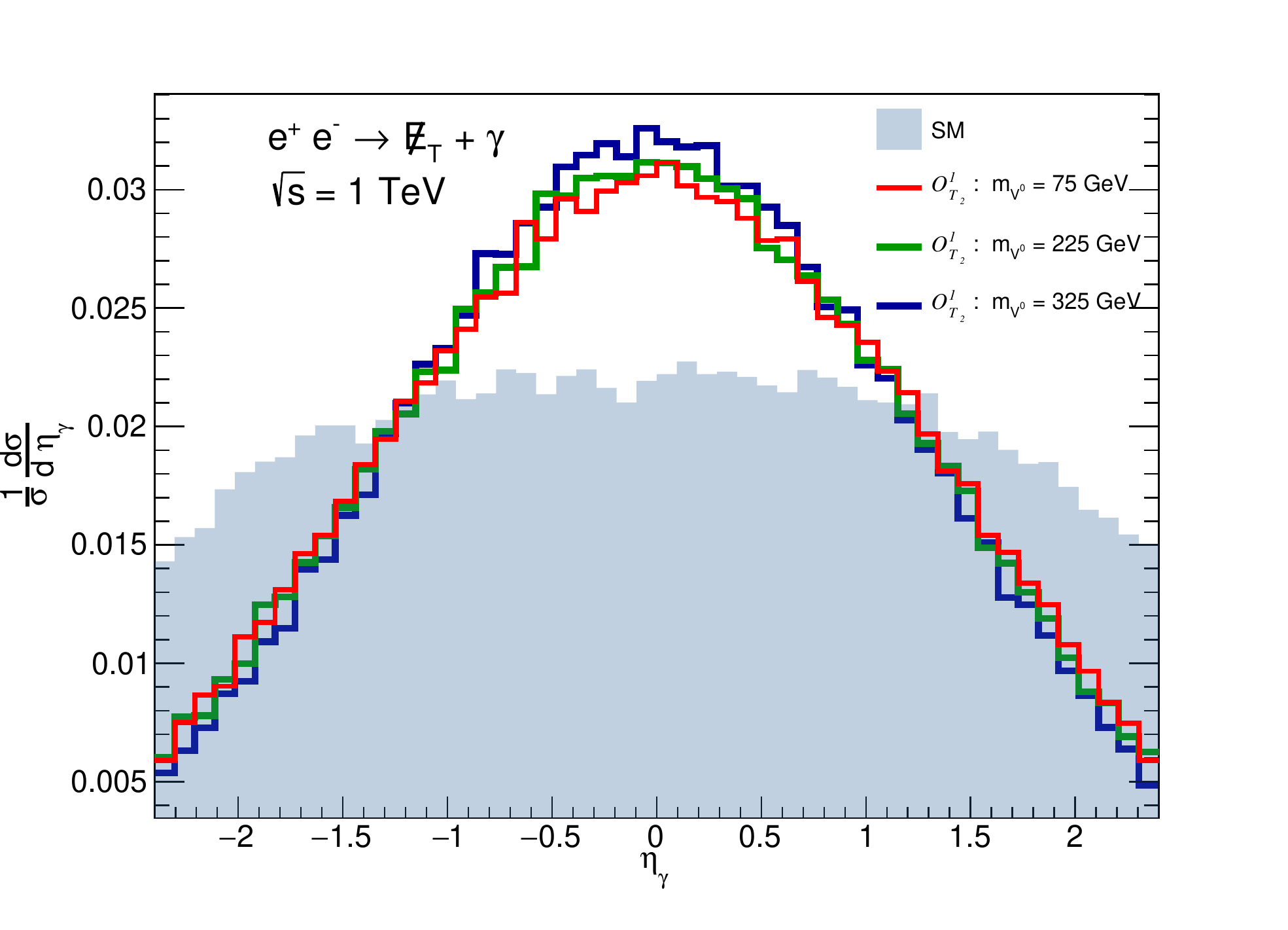}
\subcaption{\small \em{}}\label{Histo:eta_VT2}
\end{multicols}

\begin{center}	
\includegraphics[width=0.55\textwidth,clip]{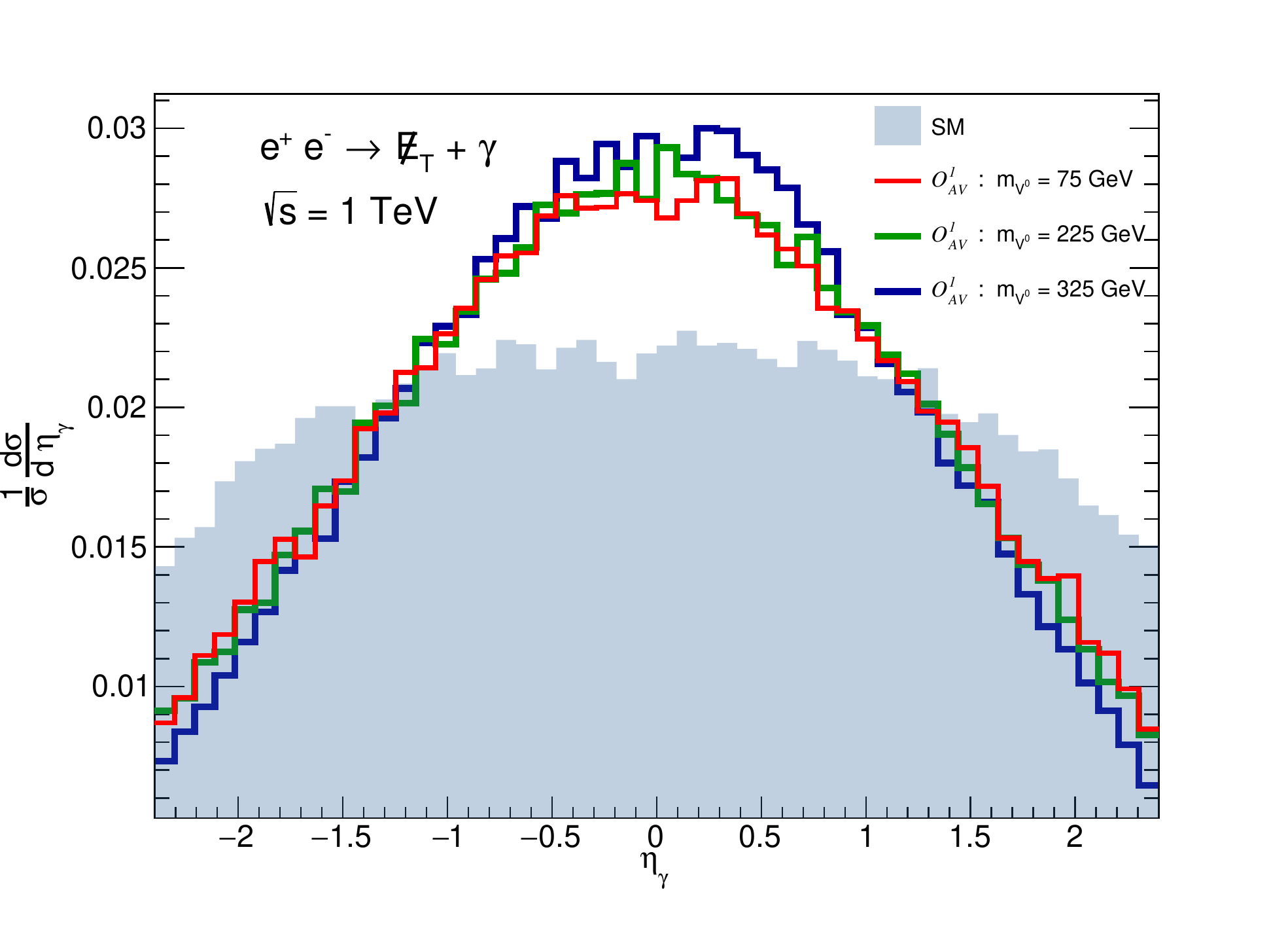}
\subcaption{\small \em{ }}
\label{Histo:eta_VAV}
\end{center}

	\caption{\small \em{Normalized 1-dimensonal differential  cross-sections with respect to $\eta_{\gamma}$ corresponding to the  SM processes and   those induced by  lepto-philic operators  at the three representative values of DM masses: 75, 225 and 325 GeV.}}
\label{fig:distEta}
\end{figure*}
\begin{figure*}
\centering
\begin{multicols}{2}
\includegraphics[width=0.49\textwidth,clip]{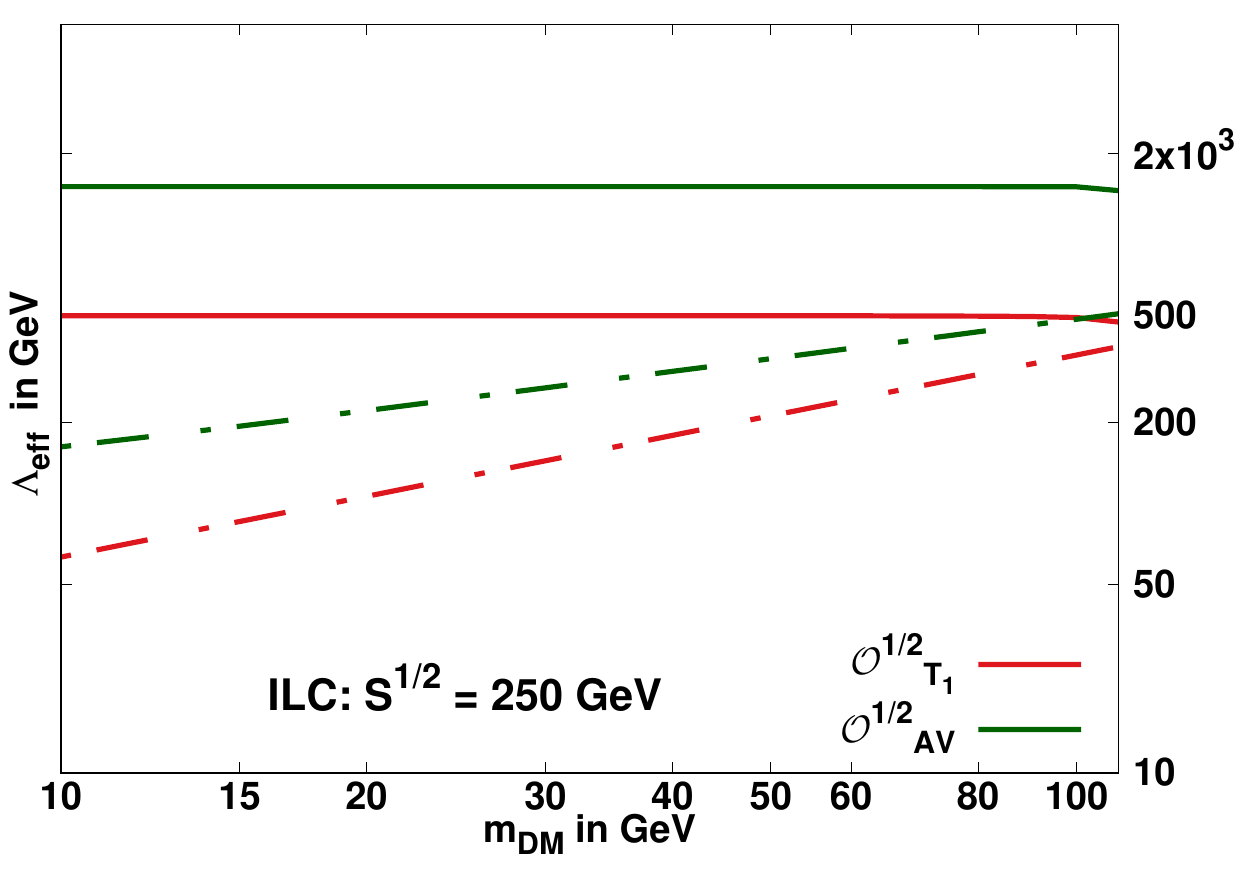}
\subcaption{\small \em{Fermionic DM }}\label{ChiSqrFDM250}
\includegraphics[width=0.49\textwidth,clip]{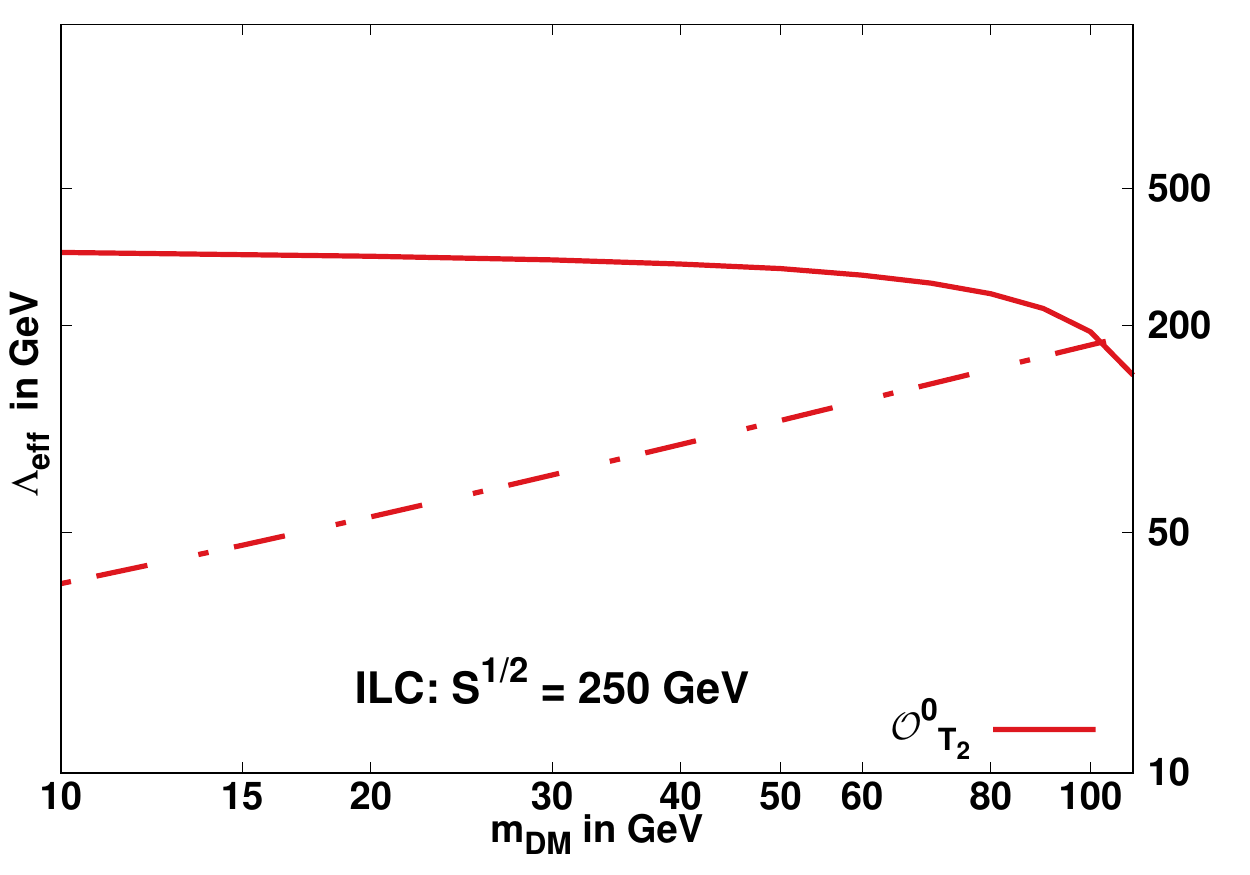}
\subcaption{\small \em{Scalar DM }}\label{ChiSqrSDM250}
\end{multicols}
\begin{center}	
\includegraphics[width=0.5\textwidth,clip]{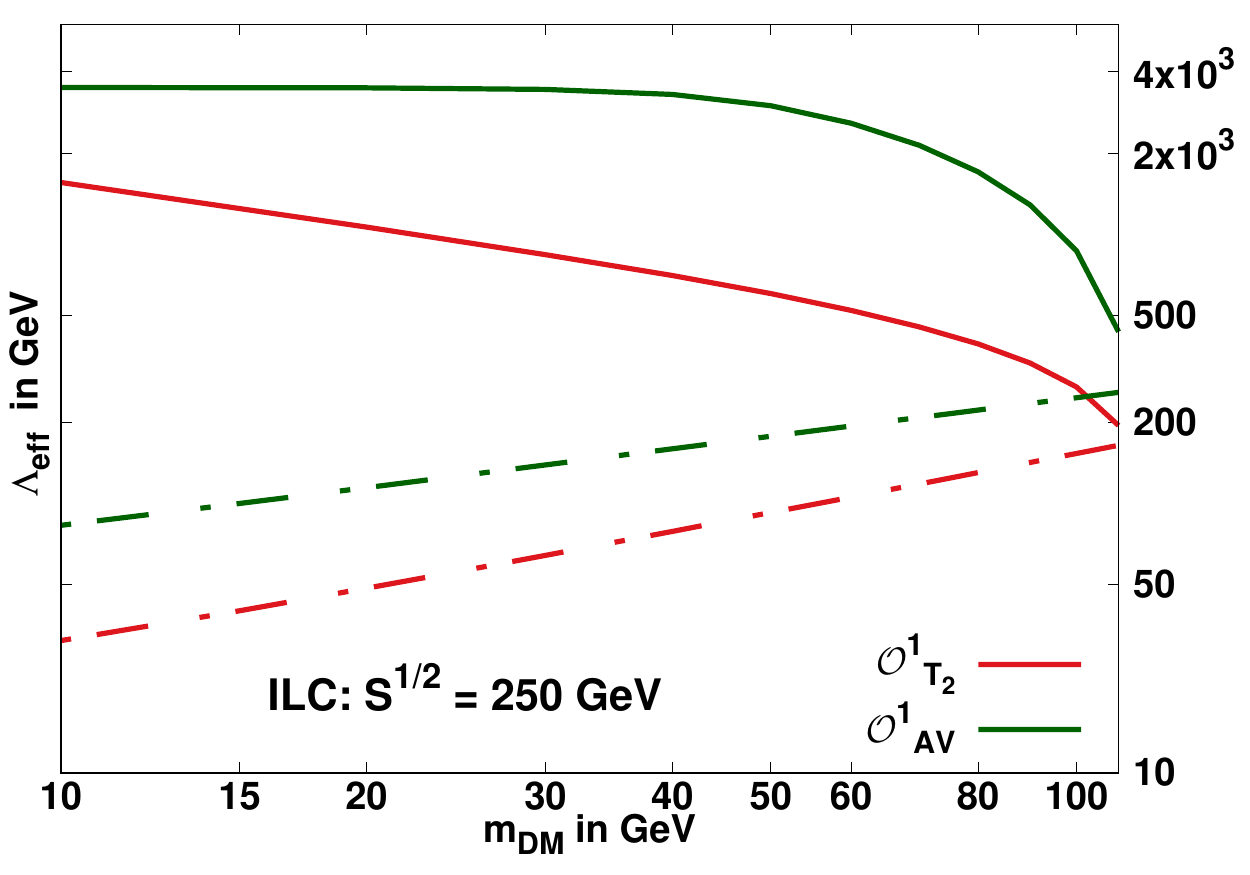}
\subcaption{\small \em{Vector DM }}
\label{ChiSqrVDM250}
\end{center}

	\caption{\small \em{Solid lines depict $3\sigma$ with 99.73 \% C.L. contours in the $m_{DM}-\Lambda_{\rm eff}$  plane from the ${{\cal X}}^2$ analyses of the $e^+e^-\to \slash\!\!\! \!E_T +\gamma$ signature at the proposed ILC  designed for $\sqrt{s}$ = 250 GeV with an integrated luminosity 250 fb$^{-1}$.
 The region below the solid lines corresponding to the respective contour is accessible for discovery with $\ge$ 99.73\% C.L.  The regions below the dashed lines corresponding to respective operators satisfy  the relic density constraint $\Omega_{\rm DM}h^2 \le$ $ 0.1198 \pm 0.0012$.}}
\label{fig:chisq250}
\end{figure*}
\begin{figure*}
\centering
\begin{multicols}{2}
\includegraphics[width=0.49\textwidth,clip]{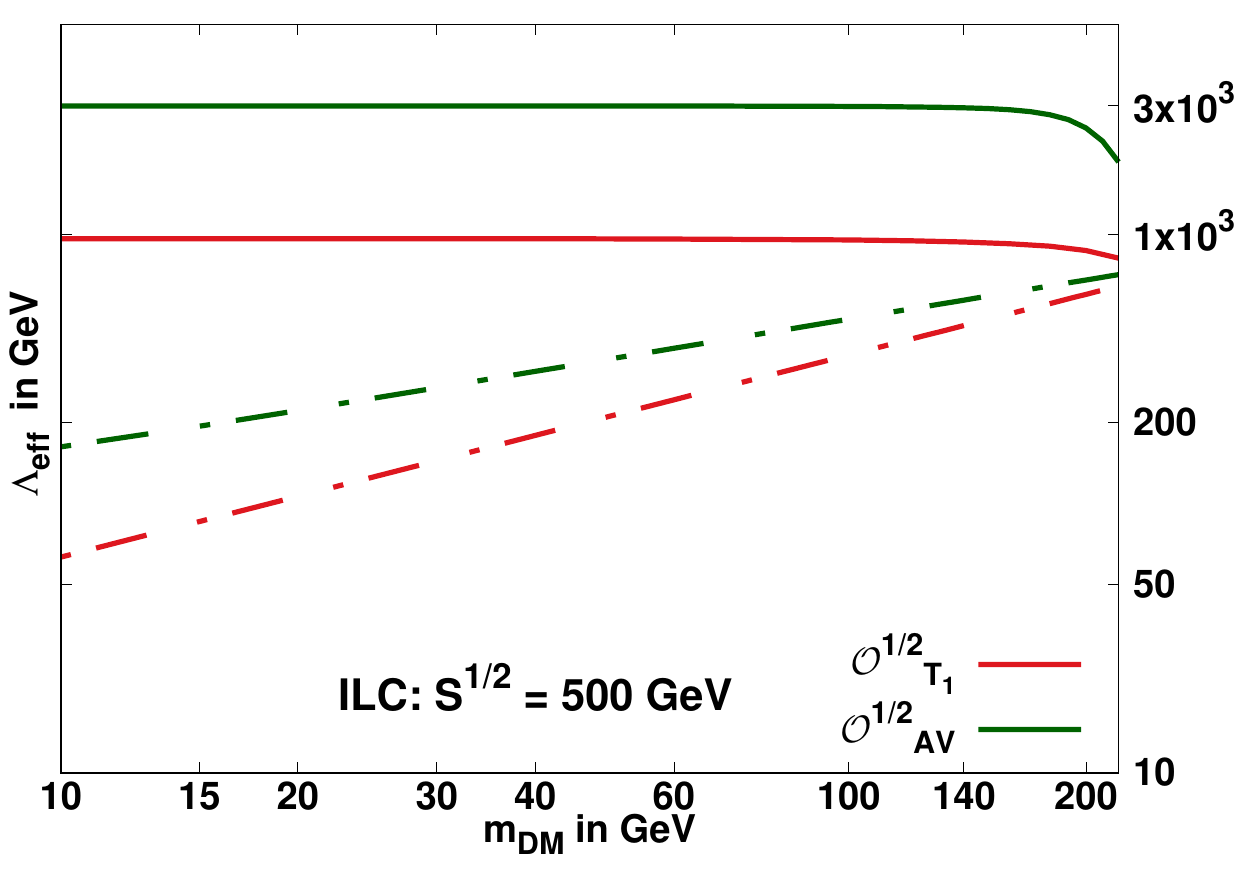}
\subcaption{\small \em{Fermionic DM }}\label{ChiSqrFDM500}
\includegraphics[width=0.49\textwidth,clip]{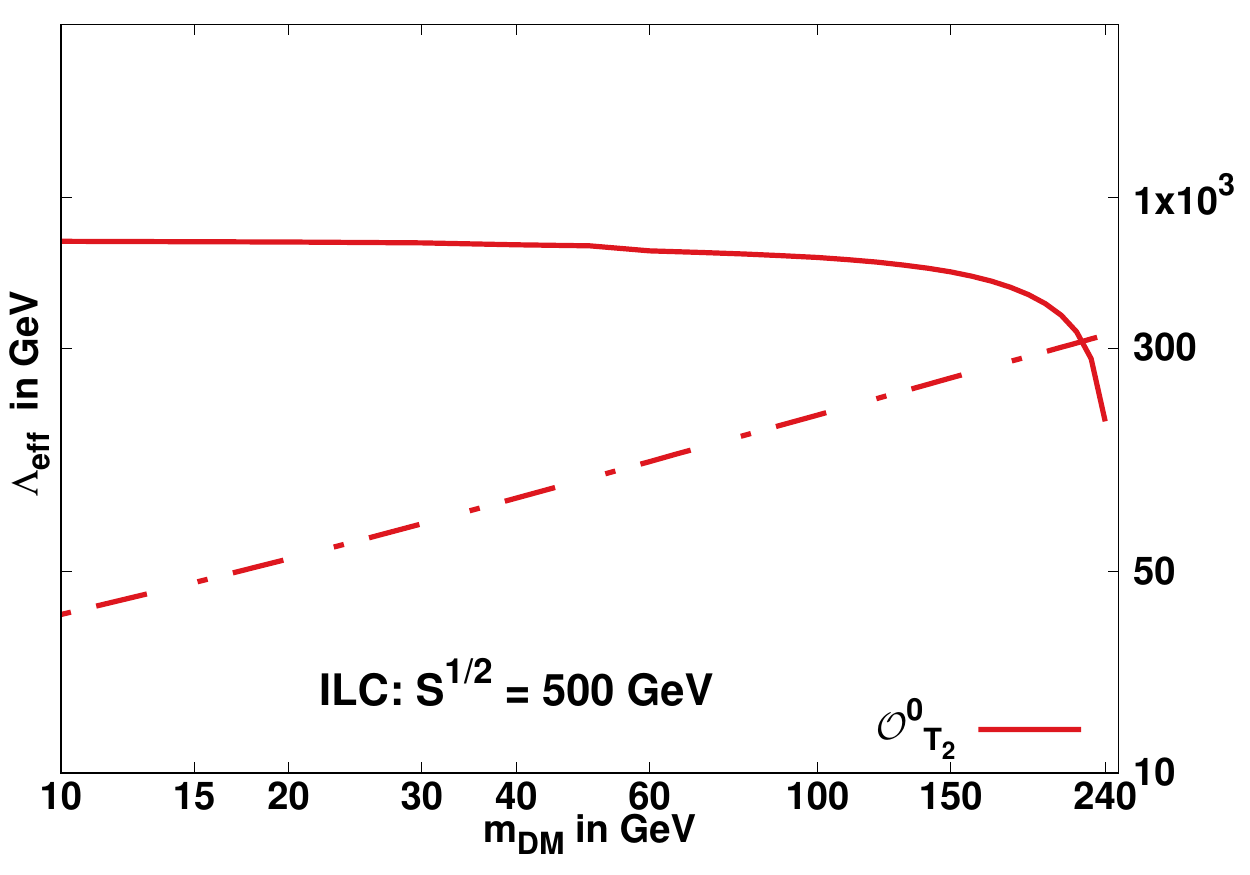}
\subcaption{\small \em{Scalar DM }}\label{ChiSqrSDM500}
\end{multicols}
\begin{center}	
\includegraphics[width=0.5\textwidth,clip]{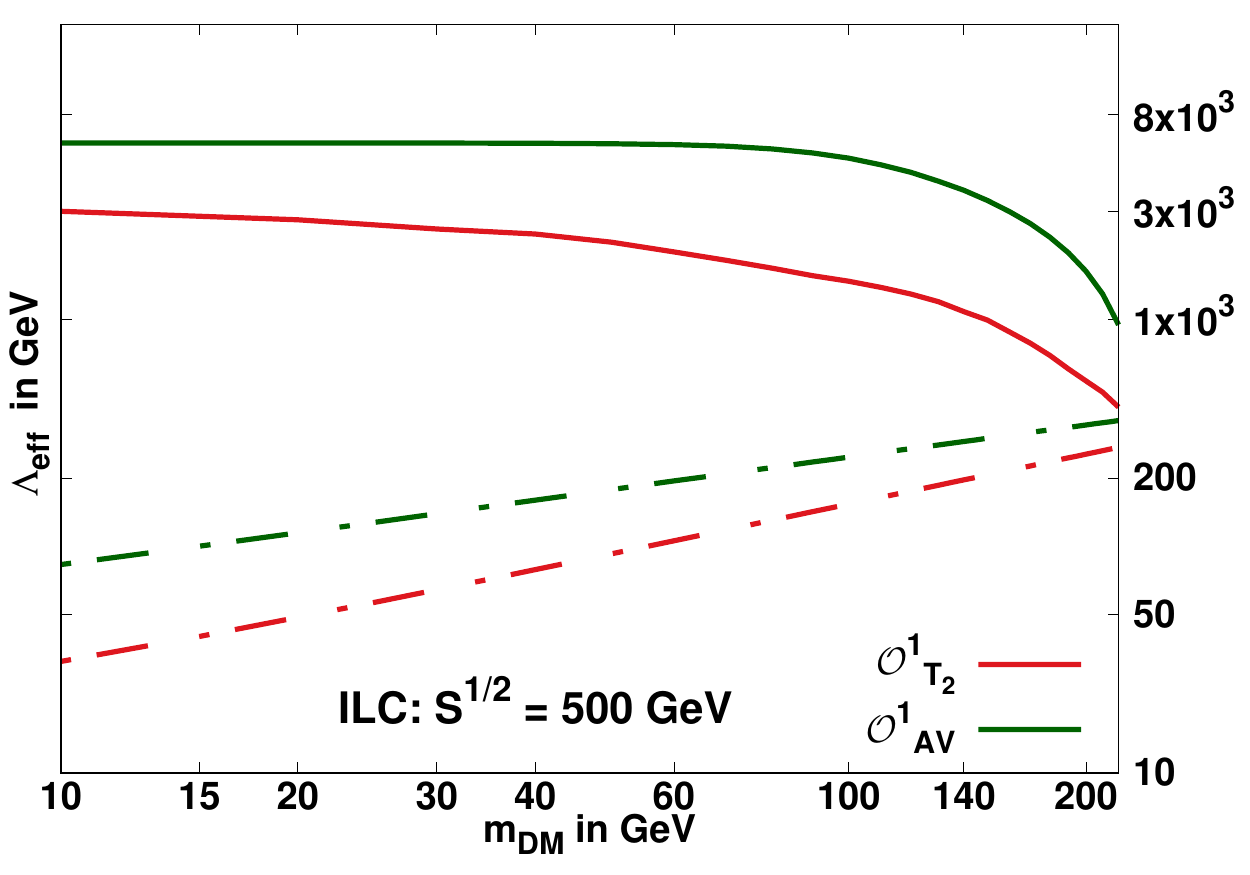}
\subcaption{\small \em{Vector DM }}
\label{ChiSqrVDM500}
\end{center}

	\caption{\small \em{Solid lines depict $3\sigma$ with 99.73 \% C.L. contours in the $m_{DM}-\Lambda_{\rm eff}$  plane from the ${{\cal X}}^2$ analyses of the $e^+e^-\to \slash\!\!\! \!E_T +\gamma$ signature at the proposed ILC  designed for $\sqrt{s}$ = 500 GeV with an integrated luminosity 500 fb$^{-1}$.
 The region below the solid lines corresponding to the respective contour is accessible for discovery with $\ge$ 99.73\% C.L.  The regions below the dashed lines corresponding to respective operators satisfy  the relic density constraint $\Omega_{\rm DM}h^2 \le$ $ 0.1198 \pm 0.0012$.}}
\label{fig:chisq500}
\end{figure*}
\begin{figure*}
\centering
\begin{multicols}{2}
\includegraphics[width=0.49\textwidth,clip]{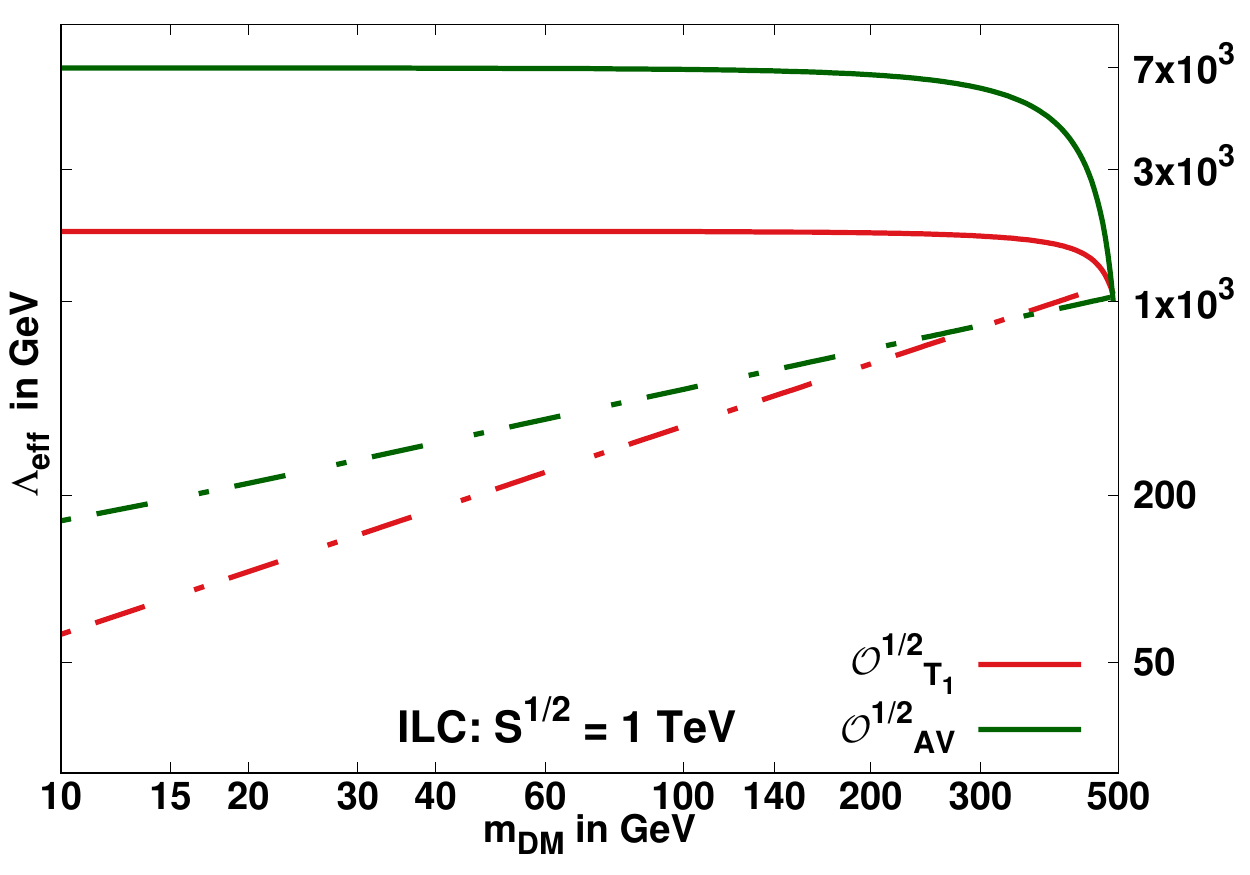}
\subcaption{\small \em{Fermionic DM }}\label{ChiSqrFDM1000}
\includegraphics[width=0.49\textwidth,clip]{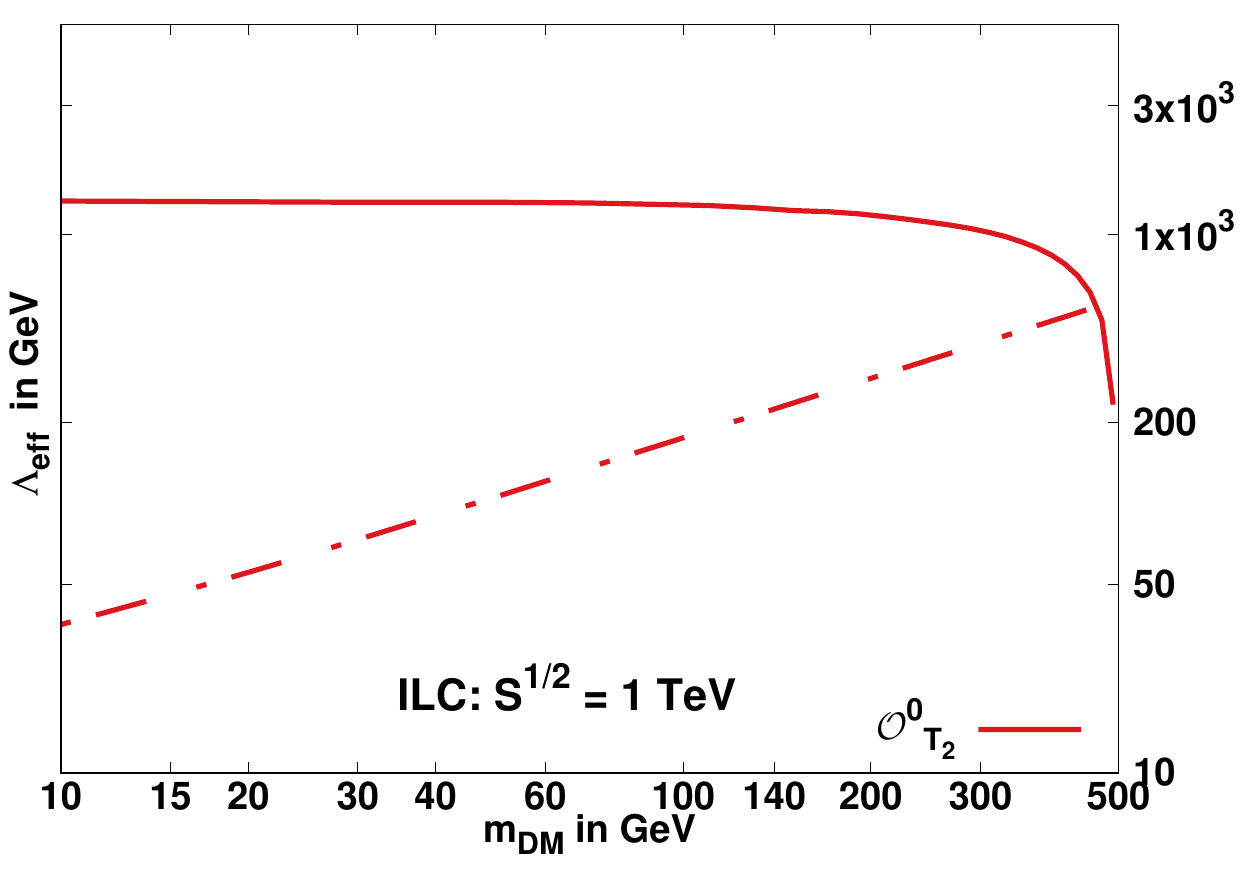}
\subcaption{\small \em{Scalar DM }}\label{ChiSqrSDM1000}
\end{multicols}
\begin{center}	
\includegraphics[width=0.5\textwidth,clip]{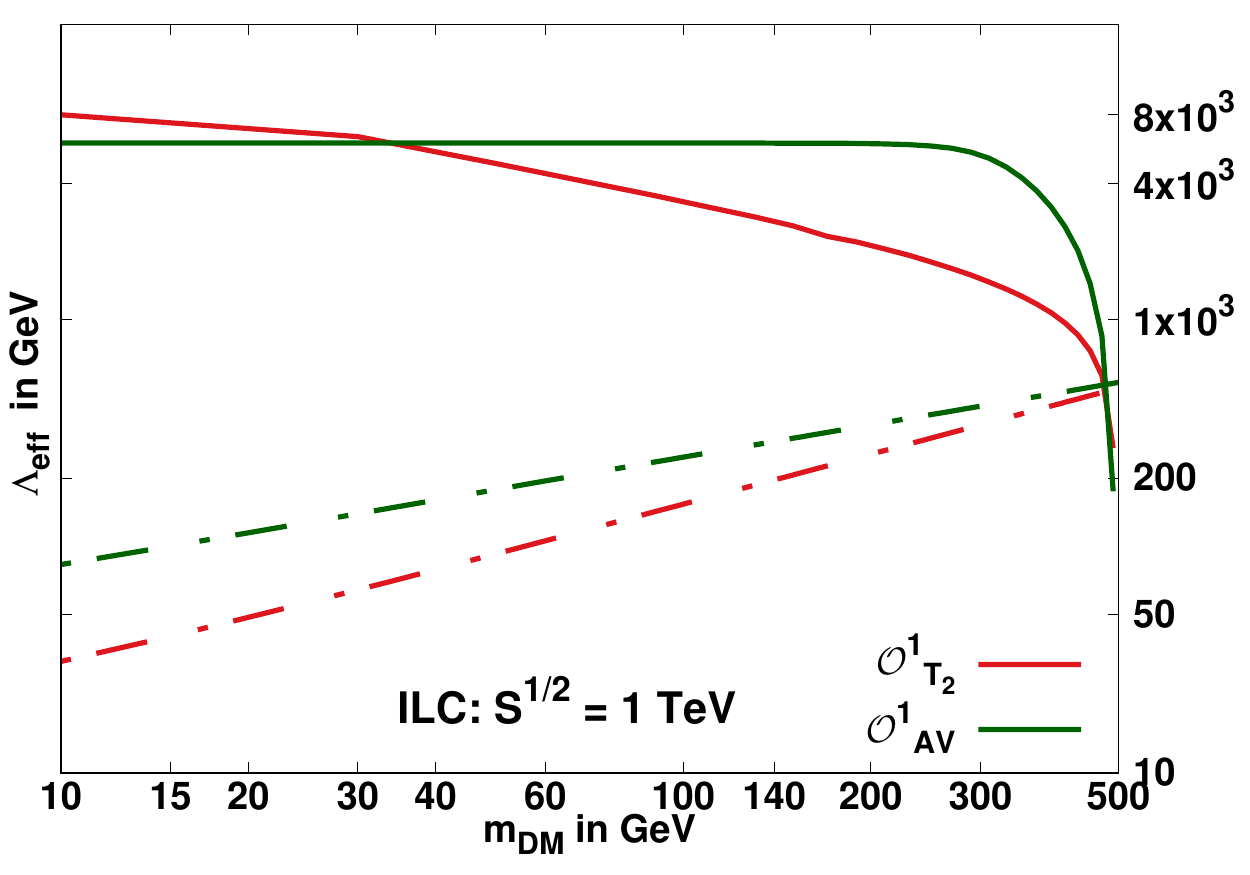}
\subcaption{\small \em{Vector DM }}
\label{ChiSqrVDM1000}
\end{center}

	\caption{\small \em{Solid lines depict $3\sigma$ with 99.73 \% C.L. contours in the $m_{DM}-\Lambda_{\rm eff}$  plane from the ${{\cal X}}^2$ analyses of the $e^+e^-\to \slash\!\!\! \!E_T +\gamma$ signature at the proposed ILC  designed for $\sqrt{s}$ = 1 TeV with an integrated luminosity 1 ab$^{-1}$.
 The region below the solid lines corresponding to the respective contour is accessible for discovery with $\ge$ 99.73\% C.L.  The regions below the dashed lines corresponding to respective operators satisfy  the relic density constraint $\Omega_{\rm DM}h^2 \le$ $0.1198 \pm 0.0012$.}}
\label{fig:chisq1000}
\end{figure*}

In this subsection we study the DM pair production processes accompanied by an  on-shell photon at the proposed ILC   
for the DM mass range  $\sim$ 50 - 500 GeV: (a) $e^+\,e^-\,\rightarrow \,\chi^0\,\bar{\chi^0}\,\gamma $,  (b) $e^+\,e^-\,\rightarrow \,\phi^0\,\phi^0\,\gamma $,  and (c) $e^+\,e^-\,\rightarrow \,V^0\,V^0\,\gamma $ 
as shown in Figures \ref{fig:chisq250}-\ref{fig:chisq1000}. The  dominant SM background  for $e^+e^-\to \not \!\! E_T + \gamma$  signature comes from $Z\gamma$ production process: $e^+\,e^-\,\rightarrow\,Z+\gamma \to \sum \nu_i\,\bar{\nu}_i + \gamma$. 
\begin{table*}[htb]\footnotesize
\centering
\begin{tabular*}{\textwidth}{c|@{\extracolsep{\fill}} ccccc|}\hline\hline
	&\textit{ILC-250}&\textit{ILC-500}&\textit{ILC-1000}\\
	$\sqrt{s} \left( \textit{in GeV}\right )$& 250 & 500 & 1000  \\
	$L_{int} \left( \textit{in $fb^{-1}$}\right )$ & 250 & 500 & 1000  \\
	$\sigma_{BG} \,(pb)$ &1.07 &  1.48 & 2.07 \\\hline\hline
\end{tabular*}
\caption{\small \em{Accelerator parameters as per Technical Design Report \cite{Behnke:2013lya, Behnke:2013xla}.  $\sigma_{BG}$ is the background cross section for $e^-\,e^+\,\rightarrow\,\sum \nu_i\,\bar{\nu}_i\,\gamma$ process computed using the  selection cuts defined in section \ref{basiccuts}}}
\label{table:accelparam}
\end{table*}

\par The analyses for the background and the signal processes corresponding to the accelerator parameters as conceived  in the {\it Technical Design Report for ILC} \cite{Behnke:2013lya, Behnke:2013xla} given in Table \ref{table:accelparam} are performed by simulating SM backgrounds and the DM signatures using Madgraph \cite{Alwall:2014hca}, MadAnalysis 5 \cite{Conte:2012fm} and  the model file generated by FeynRules \cite{Alloul:2013bka}. We impose the following cuts to reduce the backgrounds for the DM pair production in association with mono-photon:
\begin{itemize}
 \item[$\bullet$]  Transverse momentum of photon $p_{T_{\gamma}} \geq$ 10 GeV,  
 \item[$\bullet$]   Pseudo-rapidity of photon is restricted as $\left\vert\eta_\gamma\right\vert\leq$ 2.5,
 \item[$\bullet$]   dis-allowed recoil photon energy against on-shell $Z$ \\ 
$ \frac{2\,E_\gamma}{\sqrt{s}}$\ \ \  $\not\!\epsilon\ \ \ \left[0.8,0.9\right]$,\   $\left[0.95,0.98\right]$ and  $\left[0.98,0.99\right]$    for $\sqrt{s}$ = 250 GeV, 500 GeV and 1 TeV respectively.
\label{basiccuts}
\end{itemize}
\begin{table*}[ht]\footnotesize
\begin{center}
\begin{tabular}{c||c|c||c|c}
    \hline\hline
&\multicolumn{2}{c||}{\bf \underline{Unpolarised}}&\multicolumn{2}{c}{\bf \underline{Polarised}}\\
$\sqrt{s}$ in GeV&\multicolumn{2}{c||}{500}&\multicolumn{2}{c}{500}\\
$ L$ in fb$^{-1}$&\multicolumn{2}{c||}{500}&\multicolumn{2}{c}{500}\\
$\left(P_{e^-},\, P_{e^+}\right)$&\multicolumn{2}{c||}{(0,\,0)}&\multicolumn{2}{c}{(0.8,\ - 0.3)}\\
$m_{DM}$ in GeV  &$75$&$225$&$75$&$225$\\
    \hline
&&&&\\
$\mathcal{ O}^{1/2}_{T_1} $&$956.1$&$766.4$&$1135.7$&$948.0$\\
&&&&\\
$\mathcal{ O}^{1/2}_{\rm AV}$&$2994.4$&$1629.4$&$2998.6$&$2345.5$\\ 
&&&&\\
$\mathcal{ O}^{0}_{T_2}$&$461.8$&$319.1$&$767.8$&$373.2$\\
&&&&\\
$\mathcal{ O}^{1}_{T_2}$&$1751.4$&$361.8$&$1651.2$&$444.3$\\ 
&&&&\\
$\mathcal{ O}^{1}_{\rm AV}$&$5718.0$&$777.3$&$5976.2$&$1129.8$\\ 
&&&&\\ \hline\hline
\end{tabular}
\end{center}
\caption{\small \em{Estimation of $3 \sigma$ reach of the cut-off $\Lambda_{\rm eff}$ in GeV from ${\cal X}^2$ analysis for two representative values of DM mass $75$ and $225$ GeV at proposed ILC for $\sqrt{s}=500$ GeV with an integrated luminosity $500\ fb^{-1}$ for unpolarised and polarised initial beams.}}
\label{table:UnpolandPol}
\end{table*}

\par The shape profiles corresponding to the mono-photon with missing energy processes can be studied in terms of the kinematic observables $p_{{}_{T \gamma}}$ and $\eta_\gamma$ as they are found to be most sensitive. We generate the normalized one dimensional distributions for the SM background processes and signals induced by the relevant operators. To study the dependence on DM mass, we plot the normalized differential cross-sections in figures \ref{fig:distPT} \& \ref{fig:distEta} for three representative values of DM mass $75,\ 225$ and $325$ GeV at center of mass energy $\sqrt{s}=1$ TeV and an integrated luminosity 1 $ab^{-1}$.
 
 The sensitivity of $\Lambda_{\rm eff}$ with respect to DM mass is enhanced by computing the ${\cal X}^2$ with the double differential distributions of  kinematic observables  $p_{T_\gamma}$ and $\eta_\gamma$ corresponding to  the background and signal processes for  (i) 50 GeV $\le m_{\rm DM} \le$ 125 GeV  at $\sqrt{s} $ = 250 GeV and an integrated luminosity of 250 fb$^{-1}$, (ii)  100 GeV $\le m_{\rm DM} \le$ 250 GeV  at $\sqrt{s} $ = 500 GeV and an integrated luminosity of 500 fb$^{-1}$ and (iii)  100 GeV $\le m_{\rm DM} \le$ 500 GeV at $\sqrt{s} $ = 1 TeV and an integrated luminosity of 1 ab$^{-1}$.
The ${{\cal X}}^2$ is   defined as

\begin{eqnarray}
	{{\cal X}}^2&\equiv&{{\cal X}}^2 \left(m_{\rm DM},\, \frac{\alpha_i}{\Lambda_{\rm  eff}^n} \right)\nn\\
	&&\hskip -1cm =\sum_{j=1}^{n_1}\sum_{i=1}^{n_2} \left [ \frac{\frac{\Delta N_{ij}^{NP}}{\left(\Delta p_{T_\gamma}\right)_i\, \left(\Delta \eta_\gamma\right)_j}}{\sqrt{ \frac{\Delta N_{ij}^{SM+NP}} {\left(\Delta p_{T_\gamma}\right)_i\, \left(\Delta \eta_\gamma\right)_j} +\delta_{\rm sys}^2\left\{ \frac{\Delta N_{ij}^{SM+NP}}{\left(\Delta p_{T_\gamma}\right)_i\, \left(\Delta \eta_\gamma\right)_j}\right\}^2} }\right ]^2\nonumber\\
\end{eqnarray}

where $\Delta N_{ij}^{NP}$ and $\Delta N_{ij}^{SM+NP}$  are the number of New Physics 
 and  total  differential events respectively  in the two dimensional   $\left[\left(\Delta p_{T_\gamma}\right)_i-\left(\Delta \eta_\gamma\right)_j\right]^{\rm th}$ grid. Here $\delta_{\rm sys}$ represents
 the total systematic error in the measurement.
 \par Adopting a conservative value for the systematic error to be $1\%$ and using the collider parameters given in Table \ref{table:accelparam}, we simulate the two-dimension differential distributions to calculate the ${\cal X}^2$. In Figs. \ref{fig:chisq250} - \ref{fig:chisq1000} we have plotted the $3\sigma$ contours at $99.73\%$ C.L in the $m_{DM}-\Lambda_{\rm eff}$ plane corresponding to $\sqrt{s}=250$ GeV, 500 GeV and 1 TeV respectively for the effective operators satisfying the perturbative unitarity.
 
\par The sensitivity of mono-photon searches can be improved by considering the polarised initial beams \cite{Bartels:2012ex,Yu:2013aca}. For an illustrative purpose, we consider $+80\ \%$ polarised $e^-$ and $-30\ \%$ polarised $e^+$ initial beams. In Table \ref{table:UnpolandPol} we show the $3 \sigma$ reach of the cut-off $\Lambda_{\rm eff}$ from ${\cal X}^2$ analysis for two representative values of DM mass $75$ and $225$ GeV at proposed ILC for $\sqrt{s}=500$ GeV with an integrated luminosity $500\ fb^{-1}$ for unpolarised and polarised initial beams and find the improvement in the $\Lambda_{\rm eff}$ sensitivity for the polarised beams.

\section{Summary and Results}
\label{sec:summary}
In this article we have studied the DM phenomenology in an effective field theory frame work. We considered SM gauge-invariant contact interactions upto dimension 8 between the dark matter particles and the leptons. In order to ensure invariance of SM gauge symmetry at all energy scales, we have restricted ourselves to self conjugate DM particles namely a Majorana fermion, a real scalar or a real vector. We estimated their contribution to the relic density and obtained constraints on the parameters of the theory from the observed relic density $\Omega_{DM} h^2=0.1198 \pm 0.0012$. Indirect detection data from FermiLAT puts a lower limit on the allowed DM mass. The data also puts severe constraints on the twist-2 ${\cal O}^{1/2}_{T_1}$ operator for the fermionic DM and scalar ${\cal O}^0_S$ operator for the scalar DM.
\par Analysis of the existing LEP data in \ref{subsec:LEPCons}. disallows the phenomenologically interesting DM mass range $\le 50$ GeV except for the ${\cal O}^{1/2}_{AV}$ operator. We then performed ${\cal X}^2$-analysis for the pair production of DM particles at the proposed ILC for DM mass range $\sim 50-500$ GeV for the relevant operators discussed in the Table \ref{table:accelparam} We find that in the $m_{DM}-\Lambda_{\rm eff}$ region allowed by the relic density and indirect detection data, higher sensitivity can be obtained from the dominant mono-photon signal at the proposed ILC particularly for the twist-2 operators.

\section*{Note added}
 For the low mass DM, our attention was drawn by the referee to the fact that in addition to on-shell $Z$ production at LEP, the future FCC-ee and CEPC will be veritable sources of $Z$s producing  Tera $Z$s. This may result in competitive constraints \cite{Liu:2017zdh} on the twist-2 operators with covariant derivatives compared to the ISR and FSR processes considered from ILC.

\vskip 5mm

\acknowledgments
We thank Sukanta Dutta for discussions and his initial participation in this work. HB thanks  Mihoko Nojiri and Mamta Dahiya for suggestions. HB acknowledges the CSIR-JRF fellowship and support from CSIR grant  03(1340)/ 15/ EMR-II.

\appendix
\begin{center}
{\bf \Large Appendix}
\end{center}
\section{Annihilation cross-sections}
\label{AnnihilationCrosssection}

\par Annihilation cross-sections for the operators given in Eqs. \eqref{Op_FS} - \eqref{Op_VAV} are given respectively as
\begin{eqnarray}
\sigma_{S}^{\rm ann} \left\vert \vec  v\right\vert \left(\chi^0\,\bar{\chi^0}\to l^+l^-\right)&=&
 \frac{1}{8\pi}\ 
\frac{{\alpha^{\chi^0}_S}^2}{\Lambda_{\rm eff}^8}\, \, m^4_{\chi^0}\ m_l^2\  \left[ 1- \frac{m_l^2}{m_{\chi^0}^2}  \right]^{3/2}\  \left\vert \vec  v\right\vert^2
 \label{ThAvLannxsecFS} \\
\sigma_{T_1}^{\rm ann} \left\vert \vec  v\right\vert \left(\chi^0\,\bar{\chi^0}\to l^+l^-\right)&=&
\frac{1}{2 \pi}\ 
\frac{{\alpha^{\chi^0}_{T_1}}^2}{\Lambda_{\rm eff}^8}\ m^6_{\chi^0}\ \sqrt{1-\frac{m_l^2}{m_{\chi^0}^2}  } \nonumber\\
&&\times\left[2+\ \frac{m_l^2}{m_{\chi^0}^2}+\left( \frac{7}{6}-\ \frac{11}{16}\, \frac{m_l^2}{m_{\chi^0}^2}-\frac{65}{48}\, \frac{m_l^4}{m_{\chi^0}^4} \right)  \left\vert \vec  v\right\vert^2 \right] \label{ThAvLannxsecFT1} \\
\sigma_{AV}^{\rm ann} \left\vert \vec  v\right\vert \left(\chi^0\,\bar{\chi^0}\to l^+l^-\right)&=&\frac{1}{2 \pi }\ 
\frac{{\alpha^{\chi^0}_{AV}}^2}{\Lambda_{\rm eff}^4}\ m^2_l\ \sqrt{1-\frac{m_l^2}{m_{\chi^0}^2}  } \left[1+ \left( \frac{1}{3}\frac{m_{\chi^0}^2}{m_l^2} -\frac{5}{6}-\frac{7}{6}\frac{m_l^2}{m_{\chi^0}^2}\right)\ \left\vert \vec  v\right\vert^2    \right] 
\label{ThAvLannxsecFAV}\nn \\
\end{eqnarray}
\begin{eqnarray}
\sigma_{S}^{\rm ann} \left\vert \vec  v\right\vert  \left(\phi^0\,{\phi^0}\to l^+l^-\right)&=& \frac{1}{4 \pi}\ 
\frac{{\alpha^{\phi^0}_S}^2}{\Lambda_{\rm eff}^8}\, \, m^4_{\phi^0}\ m_l^2 \sqrt{1-\frac{m_l^2}{m_{\phi^0}^2}  } \left[ 1- \frac{m_l^2}{m_{\phi^0}^2} + \left( -\frac{3}{2}+ \frac{15}{4} \frac{m_l^2}{m_{\phi^0}^2} \right)\ \left\vert \vec  v\right\vert^2  \right] \nn\\
\label{ThAvLannxsecSS}
\eea
\bea
\sigma_{T_2}^{\rm ann} \left\vert \vec  v\right\vert  \left(\phi^0\,{\phi^0}\to l^+l^-\right)&=& \frac{1}{4 \pi }\ 
\frac{{\alpha^{\phi^0}_{T_2}}^2}{\Lambda_{\rm eff}^8}\, \, m^6_{\phi^0}\  \sqrt{1-\frac{m_l^2}{m_{\phi^0}^2}} \nn\\
&&  \times \left[ \frac{m_l^2}{m_{\phi^0}^2} - \frac{m_l^4}{m_{\phi^0}^4} + \left(\frac{5}{12} \frac{m_l^2}{m^2_{\phi^0}}- \frac{13}{24}\frac{m_l^4}{m^4_{\phi^0}}  \right) \ \left\vert \vec  v\right\vert^2 \right]\label{ThAvLannxsecST2}
\eea
\bea
\sigma_{S}^{\rm ann} \left\vert \vec  v\right\vert  \left(V^0\,{V^0}\to l^+l^-\right)&=& \frac{1}{12\pi}\ 
\frac{{\alpha^{V^0}_S}^2}{\Lambda_{\rm eff}^8}\, \, m^4_{V^0}\ m_l^2\ \sqrt{1-\frac{m_l^2}{m_{V^0}^2}} \nn\\
&&\times  \left[ 1- \frac{m_l^2}{m_{V^0}^2} + \left( \frac{1}{2} + \frac{7}{4} \frac{m_l^2}{m_{V^0}^2} \right) \ \left\vert \vec  v\right\vert^2  \right]
\label{ThAvLannxsecVS}
\eea
\bea
\sigma_{T_2}^{\rm ann} \left\vert \vec  v\right\vert \left(V^0\,{V^0}\to l^+l^-\right)&=& \frac{1}{12\ \pi }\ 
\frac{{\alpha^{V^0}_{T_2}}^2}{\Lambda_{\rm eff}^8}\, \, m^6_{V^0}\ \sqrt{1-\frac{m_l^2}{m_{V^0}^2}} \nn\\
&&  \times \left[ \frac{m_l^2}{m_{V^0}^2} - \frac{m_l^4}{m_{V^0}^4} + \left(\frac{3}{4} \frac{m_l^2}{m^2_{V^0}}- \frac{7}{8} \frac{m_l^4}{m^4_{V^0}}  \right)  \ \left\vert \vec  v\right\vert^2 \right] 
\label{ThAvLannxsecVT2}
\eea
\bea
\sigma_{AV}^{\rm ann} \left\vert \vec  v\right\vert  \left(V^0\,{V^0}\to l^+l^-\right)&=&  \frac{1}{54 \pi}\ 
\frac{{\alpha^{V^0}_{AV}}^2}{\Lambda_{\rm eff}^4}\, \, m^2_{V^0}\ \sqrt{1-\frac{m_l^2}{m_{V^0}^2}} \left[ 4- 7\frac{m_l^2}{m_{V^0}^2}  \right] \ \left\vert \vec  v\right\vert^2
\label{ThAvLannxsecVAV}
\end{eqnarray}


\end{document}